\documentclass[aps,prl,twocolumn,groupedaddress,showpacs,longbibliography]{revtex4-1} %
\usepackage{graphicx}
\usepackage{subfigure}
\usepackage{relsize}
\usepackage{amsmath}
\usepackage{amsfonts}
\usepackage{amssymb}
\usepackage[dvipsnames]{xcolor}
\usepackage{color} 
\usepackage{tikz}
\usetikzlibrary{shapes}
\usepackage{float}
\onecolumngrid
\definecolor{mypink3}{rgb}{0.858, 0.188, 0.478}

\begin{document}
\title{Optical Crystals and Light-Bullets in Kerr Resonators}
\author{M. Tlidi$^{1,\dagger}$, S. S. Gopalakrishnan$^{2}$, M. Taki$^{2}$, and K. Panajotov$^{3,4}$}
\affiliation{$^1$ Universit\'{e} libre de Bruxelles (ULB), Physique des Syst\`{e}mes Dynamiques, CP231, 1050 Brussels, Belgium}
\affiliation{$^2$ Facult\'{e} des Sciences, Optique Nonlin\'{e}aire Th\'{e}orique, Universit\'{e} libre de Bruxelles (ULB), CP231, 1050 Brussels, Belgium}
\affiliation{$^3$ Department of Applied Physics and Photonics (IR-TONA), Vrije Universiteit Brussels, Pleinlaan 2, 1050 Brussels, Belgium}
\affiliation{$^4$ Institute of Solid State Physics, 72 Tzarigradsko CHaussee Blvd., 1784 Sofia, Bulgaria}
\affiliation{$^5$ Laboratoire de Physique des Lasers, Atomes et Mol\'{e}cules, CNRS UMR 8523, Universit\'{e} Lille 1 - 59655 Villeneuve d'Ascq Cedex, France}

\email{$\dagger$ mtlidi@ulb.ac.be}

\date{\today}

\begin{abstract}
Stable light bullets and clusters of them are presented in the monostable regime using the mean-field Lugiato-Lefever equation [Gopalakrishnan, Panajotov, Taki, and Tlidi,
Phys. Rev. Lett. 126, 153902 (2021)].  It is shown that three-dimensional (3D) dissipative structures occur in a  strongly nonlinear regime where
modulational instability is subcritical. We provide a detailed analysis on the formation of optical 3D crystals in both the super- and sub-critical modulational instability regimes, and we highlight their link to the formation of light bullets in diffractive and dispersive Kerr resonators. We construct bifurcation diagrams associated with the formation of optical crystals in both monostable and bistable regimes. 
An analytical study has predicted the predominance of body-centered-cubic (bcc) crystals in the intracavity field over
a large variety of other 3D solutions with less symmetry. These results have been obtained using a weakly nonlinear analysis but have never been checked numerically. We show numerically that indeed the most robust structures over other self-organized crystals are the bcc crystals. Finally, we show that light-bullets and clusters of them can occur also in a bistable regime.
 \end{abstract}
\maketitle

\onecolumngrid

\section{Introduction}
The formation of macroscopic structures, whether ordered or localized, involve nonequilibrium exchanges of energy and/or matter, and has been widely observed in many natural systems including fluid mechanics, optics, biology, ecology, and medicine  \cite{cross1993pattern,arecchi1999pattern,Staliunas03,murray2007mathematical,Akhmediev08,Tlidi5,tlidi2016nonlinear}.  Driven nonlinear optical resonators, in particular, belong to this field of research and constitutes an excellent platform for researchers, to perform experimental investigations of very rich dynamics, self-organization, and symmetry-breaking instabilities. In one-dimensional (1D) dispersive systems such as macro- or microresonators, temporal localized structures (LSs) have been experimentally evidenced (see recent overview  \cite{lugiato2018kenberberg} in the theme issue  \cite{tlidi2018dissipative}). In the frequency domain, LSs display combs.  Optical frequency combs generated by microresonators have revolutionized many fields of science and technology,  such as high-precision
spectroscopy, metrology, and photonic analog-to-digital conversion
\cite{fortier201920}.  In broad area devices where diffraction cannot be ignored,  two-dimensional (2D) confinement of light leading to the formation of localized structures has been theoretically predicted in  \cite{Scroggie1}, and experimentally realized with a possibility for applications in all-optical control of light, optical storage, and information processing  \cite{Taranenko_pra00,taranenko2001spatial,Brambilla3}. 
\begin{center}
 \begin{figure}
 \unitlength=50.0mm
\centerline{
\includegraphics[width=2.2\unitlength,height=0.8\unitlength]{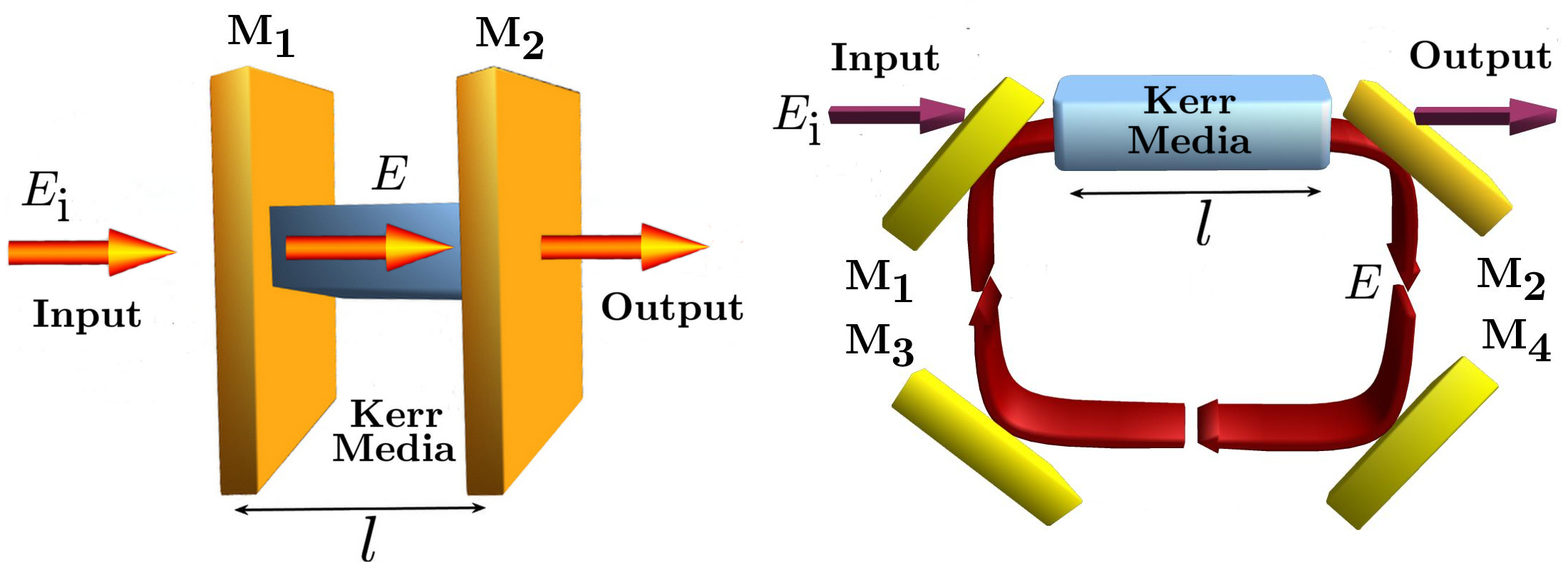}
}
\begin{picture}(0,0)
\put(0.1,0.9){\textrm{(a)}}
\put(1.1,0.9){\textrm{(b)}}
\end{picture}
 \caption{Schematic setups of optical cavities filled in with Kerr media. (a) Plane parallel cavity with length $l$ between the two mirrors 
 $\textrm{M}_1$ and $\textrm{M}_2$. (b) Ring cavity between four mirrors. The input mirros $\textrm{M}_1$ and the output mirrors $\textrm{M}_2$ have high reflectivity and are partially transmitting, while $\textrm{M}_3$ and $\textrm{M}_4$ are fully reflecting.
 }
 \label{fig:schem}
\end{figure}
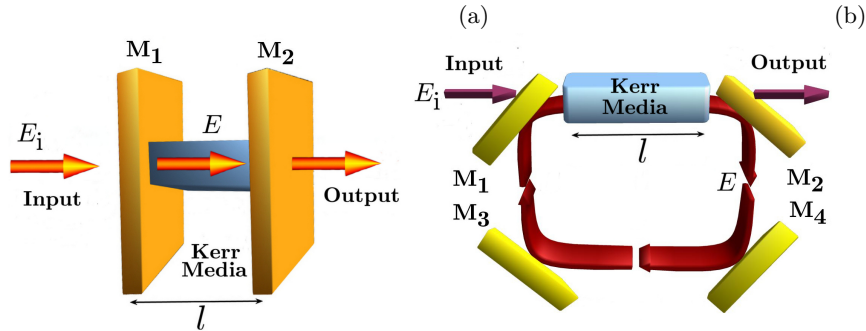
\end{center}

When both 2D diffraction and 1D dispersion have a comparable influence during light propagation in a Kerr resonator, light bullet suffers  collapse beam phenomena in the case of the 3D nonlinear Schrödinger equation  \cite{Silberberg1,Edmundson1}. By introducing additional physical effects, it is possible to avoid the collapse and to stabilise the LB formation. Several physical effect have been proposed in the literature such as Kerr cavities \cite{Tlidi4,Tlidi2,Brambilla1}. The existence of stable LBs have been reported in other systems such as in wide-aperture lasers with a saturable absorber \cite{Rozanov1,Veretenov1,Marconi1,Javaloyes1,Gurev1}, optical parametric oscillators \cite{Staliunas1,Veretenov2,Panoiu1}, second harmonic generation \cite{Tlidi3,tlidi2000three}, passively mode-locked semiconductor lasers \cite{Grelu12}, left-handed materials \cite{kockaert2006negative}, twisted waveguide arrays \cite{Torner1}, in Swif-Hohenberg equation \cite{Staliunas1,Tlidi3,Clerc}, and in the complex cubic-quintic Ginzburg--Landau equation \cite{mihalache2006stable}.
 (see recent reviews \cite{Malomed1,Mihalache14,Malomed19}).

In broad area Kerr resonators light bullets are generated. They consist of self-organized structures that travel with the group velocity of the light within the cavity.   Their stabilization is attributed to not only a balance between nonlinearity and diffraction/dispersion, but also the second balance that involves pumping or injection and dissipation or losses.  This combined action of 1D dispersion and 2D  diffraction in a Kerr resonator has revealed the existence of three-dimensional (3D) dissipative structures that can be spontaneously generated  \cite{Tlidi4,Tlidi2}. Weakly nonlinear analysis and the relative stability analysis in the neighbourhood of 3D modulational instability has shown the predominance of the body-centered-cubic (bcc) lattice structure over other periodic structures such as lamellae, face-centered-cubic, or hexagonally
packed cylinders. These analytical results however have never been checked numerically. The purpose of this paper is two fold: Firstly, to clarify the formation of optical crystals that emerge from the modulational instability, and secondly, to study the implication of the subcritical modulational instability on the formation of light bullets and clusters of them.

For this purpose, we consider optical resonators filled with a Kerr medium and coherently driven by an external injected field $E_{\textrm{i}}$.   The schematic setup of a Fabry-Perot or a ring cavity setups are shown in Fig. \ref{fig:schem}.  The transmitted part of this field interacts with the nonlinear media and suffers from nonlinearity, diffraction, chromatic dispersion, and losses. The physics of the Kerr optical resonator is best described by the paradigmatic Lugiato-Lefever equation (LLE) \cite{Lefever1}. This model consists of a damped, and driven nonlinear Schr\"{o}dinger equation, with detuning which was originally derived to describe diffractive spatial Kerr resonators. In this case, 2D diffraction ensures a coupling between different points in the transverse plane. When diffraction is neglected by using wave-guided structures such as fibers,  the inclusion of the chromatic dispersion in the dispersive resonators leads also to the temporal LLE \cite{haelterman1992dissipative}.  When 1D dispersion and 2D  diffraction have a comparable influence, the formulation of this problem leads to the generalized Lugiato-Lefever equation \cite{Tlidi2,Tlidi4}
\begin{equation}
\frac{\partial E}{\partial t} = E_\textrm{i} - ( 1 + i\delta) E + \Big(\nabla_{\perp}^2  + \frac{\partial^2}{\partial \tau^2}\Big) E + i\vert E \vert^2 E,
\label{eqn:LLE}
\end{equation}where $E=E(x,y,t,\tau)\rightarrow(\kappa/\gamma l)^{1/2}E(x,y,t,\tau)$ is the normalized slowly-varying envelope of the
electric field, with $\kappa$ being the total losses, $\gamma$ is the nonlinear coefficient, and $l$ is the cavity length. The detuning parameter $\delta=\phi/ \kappa$ is the cavity detuning parameter where $\phi$  is the linear phase shift accumulated by the intracavity field over the cavity length $l$. The injected field   $E_\textrm{i}\rightarrow \kappa(\kappa/\gamma \delta l)^{1/2} E_\textrm{i}$  is real, positive
and constant assuming a continuous wave (CW) operation.  The transverse Laplacian acting on the transverse plane $(x,y)$ is denoted by $\nabla_{\perp}^2$, and the second-order derivative term has a positive coefficient so that the cavity operates in the anomalous dispersion regime. In this case, the operator  $\nabla_{\perp}^2  + \partial^2_ {\tau }=\partial^2_{x} + \partial^2_{y}+\partial^2_ {\tau }$ is the 3D Laplacian acting in the Euclidian $(x, y, \tau )$ space. Time $t$ is the slow time describing the
evolution over successive round trips, and $\tau$ is the fast time in the
reference frame moving with the group velocity of the light within the cavity. In terms of physical
parameters, the transverse coordinate, slow, and fast times are
\begin{equation}
(x,y)\rightarrow \sqrt{\frac{l}{2 q \kappa}}(x,y), { \mbox{ } } (t,\tau) \rightarrow \Big(\frac{t_r}{\kappa}t, \sqrt{\frac{\beta_2l}{2\kappa}}\tau\Big), \nonumber \\ 
\end{equation}
where $t_r$ is the round trip time and $\beta_2$ denotes the second-order chromatic dispersion coefficient of the Kerr material. The LLE has been derived for other systems such as liquid crystals, left-handed
materials \cite{kockaert2006negative}, and photonics coupled waveguides \cite{peschel2004discrete}, whispering-gallery-mode microresonators \cite{Chembo1}. In early reports, the  LLE has been derived for a plasma driven by an external radiofrequency
field \cite{morales1974ponderomotive} and for the condensate in the presence of an applied ac field  \cite{kaup1978theory}. Due to the richness of its broad spectrum of space-time dynamical behaviors, this simple model has attracted considerable theoretical and experimental investigations during these last decades, as
witnessed by recent overviews \cite{Chembo2017theory,lugiato2018kenberberg}.

The paper is organized as follows. In section 2\ref{sec:3Dopt}, we present numerical simulations of the LLE Eq. \ref{eqn:LLE} showing indeed that the only stable optical crystals in the neighbourhood of the 2D  modulational instability are indeed the bcc structures. This result has been established theoretically in previous reports in the weakly nonlinear regime \cite{Tlidi2,Tlidi4} but never checked numerically. We construct the bifurcation diagram and we compare the results obtained by numerical simulation with these obtained through a normal form analysis. In section 3\ref{sec:lightbullet}, we consider a bistable regime where the 3D modulational instability appears subcritical. In this case, a pinning range of parameters exists where stable light-bullets and clusters of them can be generated. We construct their bifurcation diagram and we show that their domain of stability is wider than the monostable case studied recently. In addition, we obtain the stationary single light-bullet solution of the LLE Eq. \ref{eqn:LLE} by using a spherical approximation, and we compare it with a direct numerical simulation of the governing equation in section 3\ref{sec:lightbullet}. We conclude in section 4\ref{sec:discconc}.

\section{3D modulational instablity and optical crystals} \label{sec:3Dopt}

In 2D settings, numerical simulations indicate that only hexagonal structures are stable close to the modulational instability \cite{PhysRevA.46.R3609,Tlidi6}. The weakly nonlinear analysis has allowed an investigation on the existence and stability of different periodic solutions, such as hexagons and stripes. The pattern selection analysis consists of studying the stability of one pattern to perturbations favoring another pattern \cite{PhysRevA.46.R3609}. This analysis is referred to as the relative stability analysis, and has shown analytically that stripes are not stable \cite{Tlidi2}, which we confirm numerically in this study. The same analysis has been extended to  3D settings and has revealed the predominance of the body-centered-cubic (bcc) lattice structure over a variety of 3D structures in the cavity field intensity \cite{Tlidi6}. However, numerical simulations of 3D optical crystals are missing. 
The purpose of this paper is to bridge this gap and to present numerical simulations that confirm the analytical predictions obtained by the weakly nonlinear analysis.

Previous studies that have attempted to solve the 3D LLE (1) have used low-order finite-difference schemes coupled with low order Euler time stepping, which is indeed prone to numerical instabilities. This is mainly due to the fact that the LLE couples a stiff diffusion term with a strongly nonlinear term, which when discretised leads to large systems of strongly nonlinear stiff ordinary differential equations (ODEs)  \cite{Jones1,Kassam2}.
In addition finite-difference methods can sometimes lead
to spurious solutions which are non-physical  \cite{Jones1}, which
is where higher-order spectral methods come to the fore.
In the present work, the temporal discretisation is 
carried out with a fourth order exponential time differencing Runge–-Kutta method \cite{Trefethen1,Matthews1}, and the
spatial discretisation of the LLE
is done using a Fourier spectral method with periodic
boundary conditions \cite{Trefethen1,Kassam2,Saad1}. 
In the resulting discretised set of ODEs the linear term is diagonal, which is one of the main advantages 
of using a Fourier spectral method. The nonlinear term is evaluated
in physical space and then transformed to Fourier space.
A detailed analysis on these methods can be found in these
excellent books \cite{Trefethen1,Kassam2,Saad1} . 
In this study, we use a periodic domain of size
$[0, 80]^{3}$ units, which is found to be sufficient for the present study, discretised
using $128$ grid points in each direction, with a time-step
of $0.01$.

In the absence of diffraction and dispersion, the homogeneous steady state solutions of LLE, satisfying $\nabla_{\perp}^2E_s=0$, $\partial^2_ {\tau }E_s=0$, and  $\partial_ {t }E_s=0$, are given by 
$E_\textrm{i}^{2} = \vert E_s \vert^{2}[1 + (\delta - \vert E_s \vert^{2})^{2}]$. For $\delta < \sqrt{3}$, the transmitted intensity as a function of the input intensity $E_\textrm{i}^{2}$ is single-valued, whereas bistability occurs for $\delta > \sqrt{3}$. We consider small perturbations that depend on the coordinates  $(x,y,t,\tau)$ in the form of  plane waves $\exp{[i {\bf k_{\perp}} \cdot {\bf r}+ik_{\tau}\tau+\sigma t]}$. This formulation leads to the following characteristic equation
\begin{equation}
\sigma+2 \sigma+\frac{\partial I_s}{\partial I_i}+(k^2_{\perp} + k^2_{\tau})[k^2_{\perp} + k^2_{\tau}-2(2I_s-\delta)]=0
\end{equation}
where $\partial I_s/\partial I_i=1+(I_s-\delta)(I_s-2\delta)$ is the slope of the homogeneous steady states. These states undergo a modulational instability  when $\sigma=0$, and $\partial \sigma/\partial k^2=0$ with $k^2=k^2_{\perp} + k^2_{\tau}$. The threshold associated with the MI is  $E_{\textrm{ic}}^{2}=1 + (\delta - 1)^2$ for the injected field intensity. The corresponding intracavity intensity is $\vert E_\textrm{c} \vert^{2} = 1$. At this bifurcation point, the wavelength of 3D patterns is $\Lambda=2\pi/\sqrt{2-\delta}$.  When increasing the injected field above its value at the MI, there exists a finite band of Fourier modes $k$, $k^2_{\perp-} + k^2_{\tau-}<k^2_{\perp} + k^2_{\tau}< k^2_{\perp+} + k^2_{\tau+}$ with
\begin{equation}
k^2_{\perp \pm} + k^2_{\tau \pm}=2I_s-\delta \pm \sqrt{I_s^2-1},
\end{equation}
which are linearly unstable and trigger the spontaneous evolution of the intracavity field towards a self-organized optical crystal. These structures consist of regular 3D lattices of bright
spots traveling at the group velocity of light within the cavity. 

\begin{center}
 \begin{figure}
 \unitlength=50.0mm
\centerline{
\includegraphics[width=0.8\unitlength,height=0.8\unitlength]{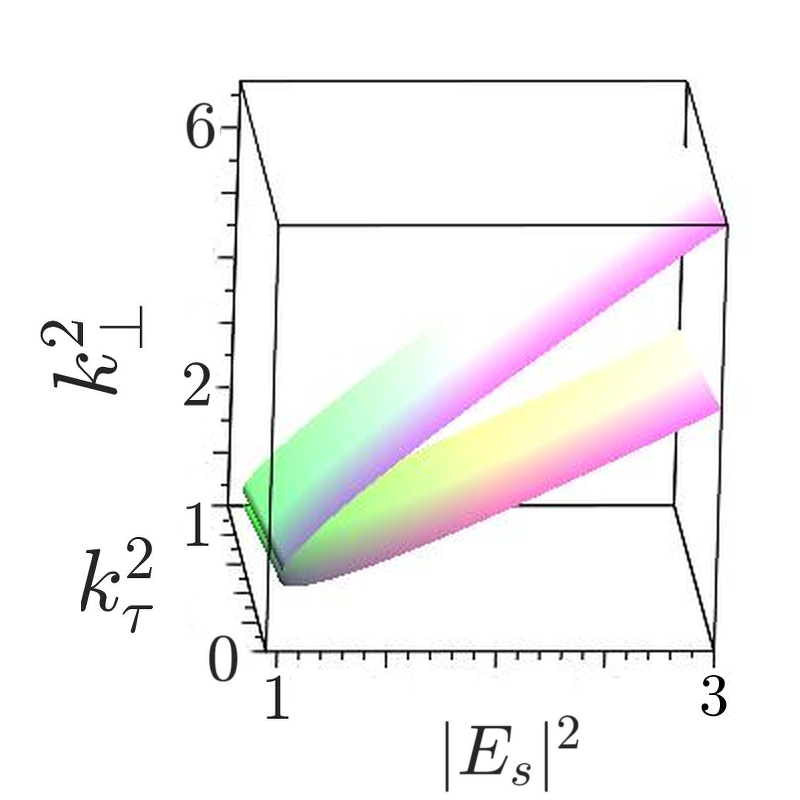}
\includegraphics[width=1.20\unitlength,height=0.8\unitlength]{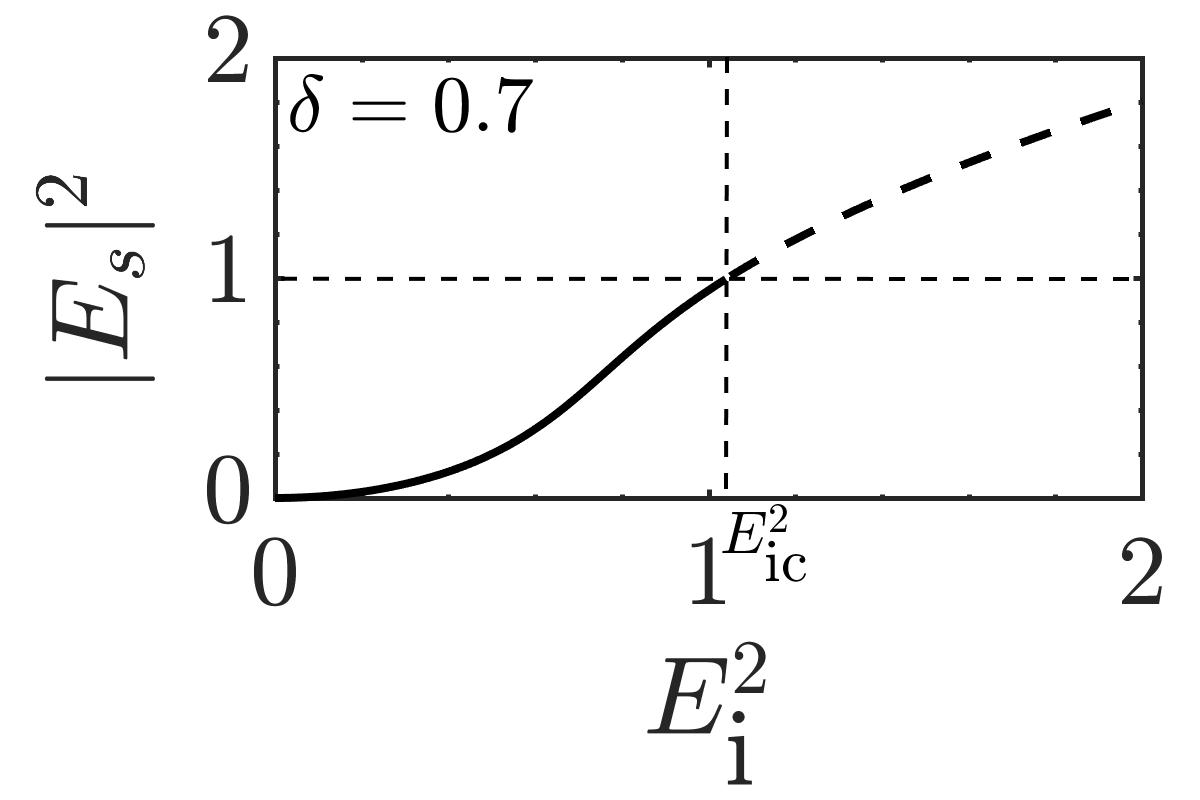}
}
\centerline{
\includegraphics[width=0.8\unitlength,height=0.8\unitlength]{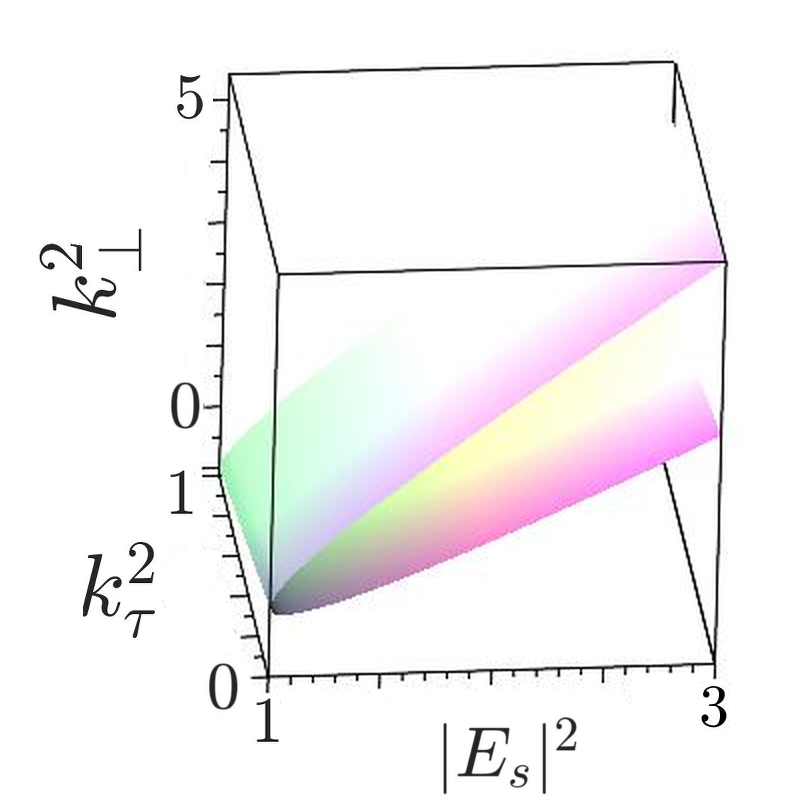}
\includegraphics[width=1.20\unitlength,height=0.8\unitlength]{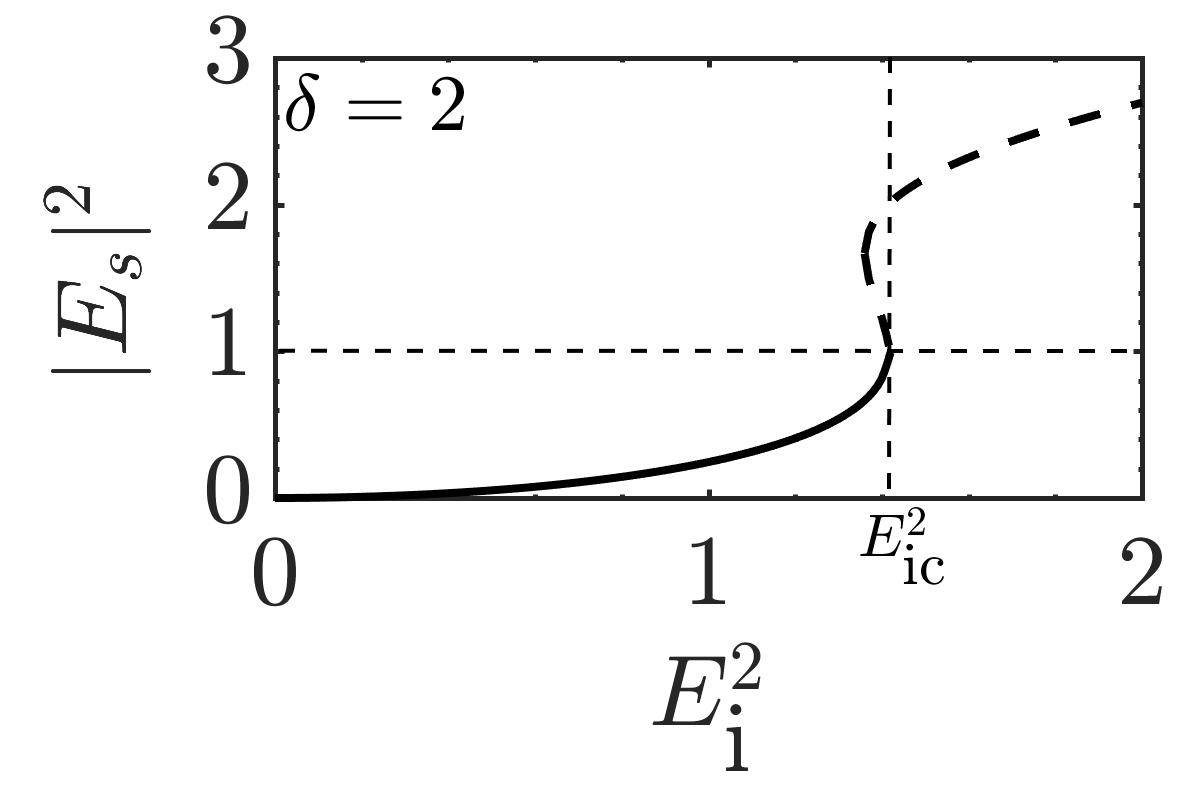}
}
\begin{picture}(0,0)
\put(0.1,1.7){\textrm{(a)}}
\put(1.0,1.7){\textrm{(b)}}
\put(0.1,0.9){\textrm{(c)}}
\put(1.0,0.9){\textrm{(d)}}
\end{picture}
 \caption{Homogeneous solution for $\delta = 0.7$ and $\delta = 2$.  (a, c) Stable and unstable 3D modes in the plane $(\vert E_s \vert^2, k^2)$. The 3D wavenumbers corresponding to 2D diffraction and 1D dispersion given by  $k^2 = k^2_{\perp} + k^2_{\tau}$, with $k^2_{\perp} = k^2_{x} + k^2_{y}$.
(b, d) Homogeneous steady states $\vert E_s \vert^2$ as a function of the input intensity $E^2_{\textrm{i}}$.  This suffers a 3D modulational  instability at  $E_\textrm{i}=E_\textrm{ic}$. Broken lines correspond to unstable solutions.  }
 \label{fig:Fig2}
\end{figure}
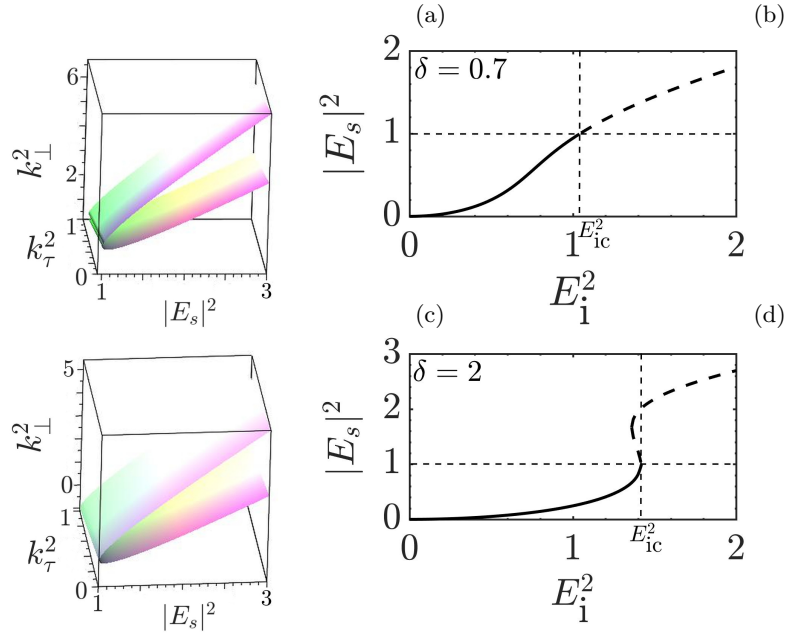
\end{center}

The marginal stability curves together with the characteristic input-output are shown in Fig. \ref{fig:Fig2} for two different values of the detuning parameter $\delta$. The number of unstable Fourier modes is much larger than in the 2D setting. These modes are arbitrarily directed in the Fourier space $(k_x,k_y,k_{\tau})$ since the system is isotropic in the Euclidean $(x,y,t,\tau)$ space. The  maximum gain or the most unstable
wave number is $k_c^2= k^2_{x} +  k^2_{y}+ k^2_{\tau}=2-\delta$. These modes form a sphere  of radius $\sqrt{2-\delta}$ in Fourier
space $(k_x,k_y,k_{\tau})$. There exists an indefinite number of modes generated with arbitrary directions. However, the nonlinear interaction allows for the generation and selection of regular crystals. Close to the MI threshold, three-dimensional periodic crystals, are approximated by a linear
superposition of $n$ pairs of opposite wave vectors $k_j$  lying on the critical sphere of radius $\sqrt{2-\delta}$ as
\begin{equation}
E(\bf{r},t) = E_s + \bf{e}\sum_{j=1}^{n} A_j \exp{(\dot{\imath}\bf{r} \cdot \bf{k_{j}})} +c.c.
\end{equation}
where  $\textrm{c.c}$ denotes the complex conjugate, and ${\bf{e}}=(2-\delta)/\delta$ is the eigenvector of the corresponding Jacobian matrix associated with the zero eigenvalue.  The 
lamellae and rhombic structures are characterized by $n=1$ and $n=2$ respectively, and the 3D hexagons or hexagonally
packed cylinders correspond to $n=3$ with $\sum_{j=1}^{3}\bf{k_j}=0 $. The face-centered-cubic (fcc) lattice and the quasiperiodic crystals are obtained for $n=4$ and  $n=5$, respectively. The body-centered-cubic (bcc) lattice corresponds to $n=6$ with the resonance conditions.

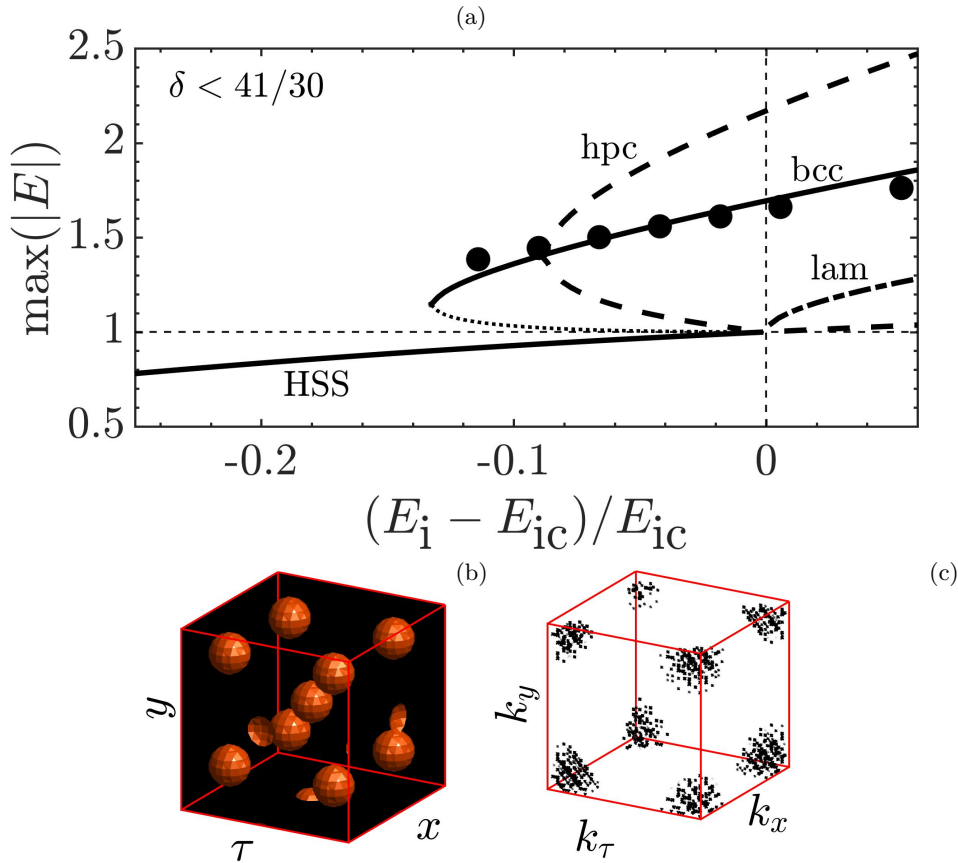
\begin{figure}
 \unitlength=60.0mm
 \centerline{
\includegraphics[width=2.167\unitlength,height=1.21\unitlength]{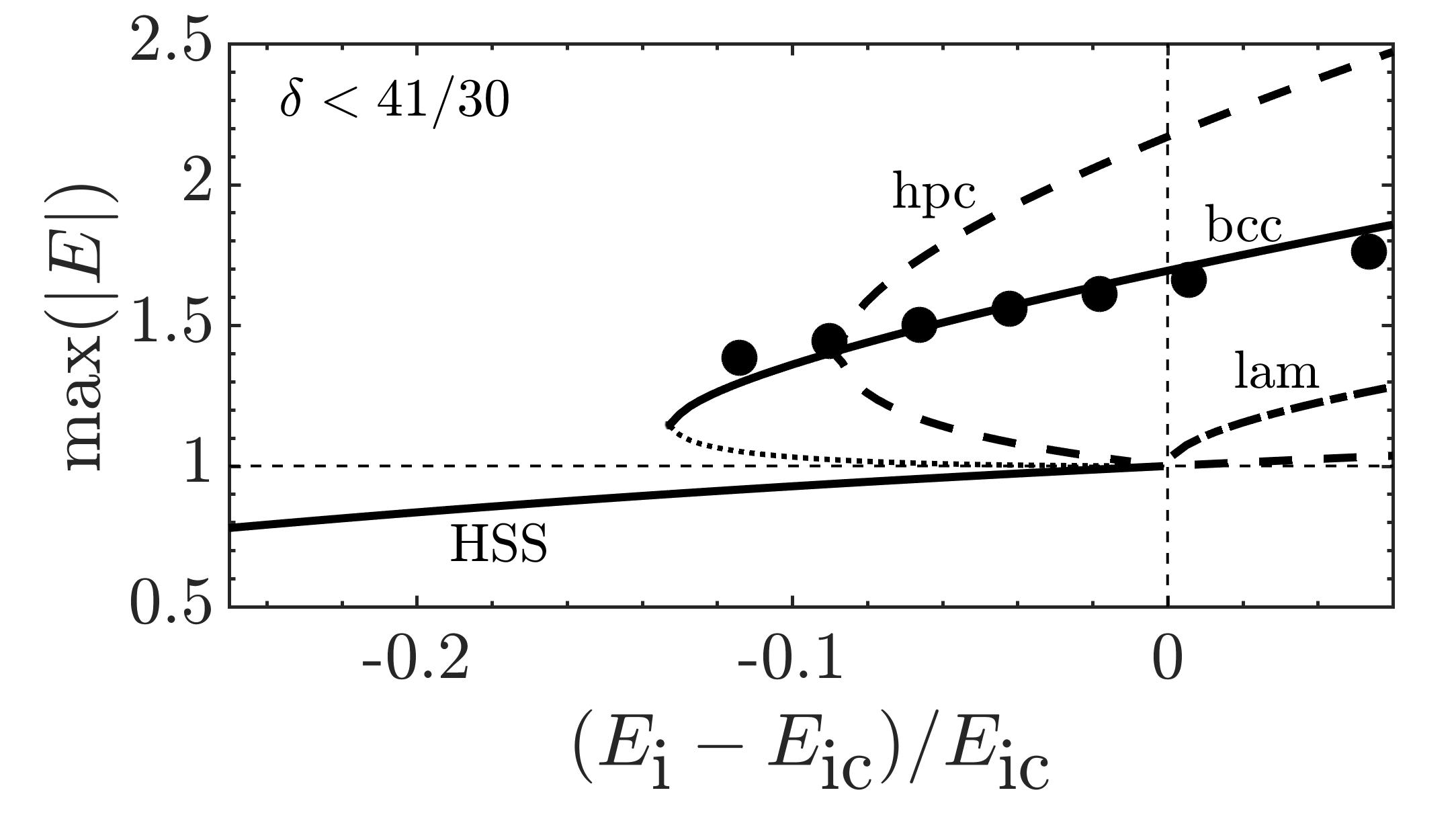}
}
 \unitlength=20.0mm
 \centerline{
\includegraphics[width=2.0\unitlength,height=2.0\unitlength]{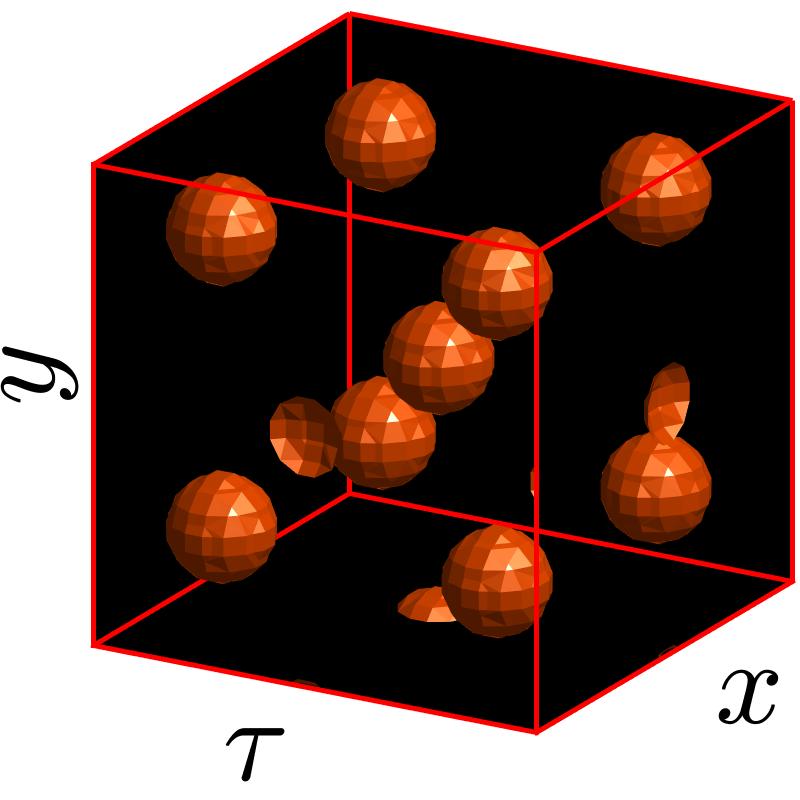}\hspace{0.5cm}
\includegraphics[width=2.0\unitlength,height=2.0\unitlength]{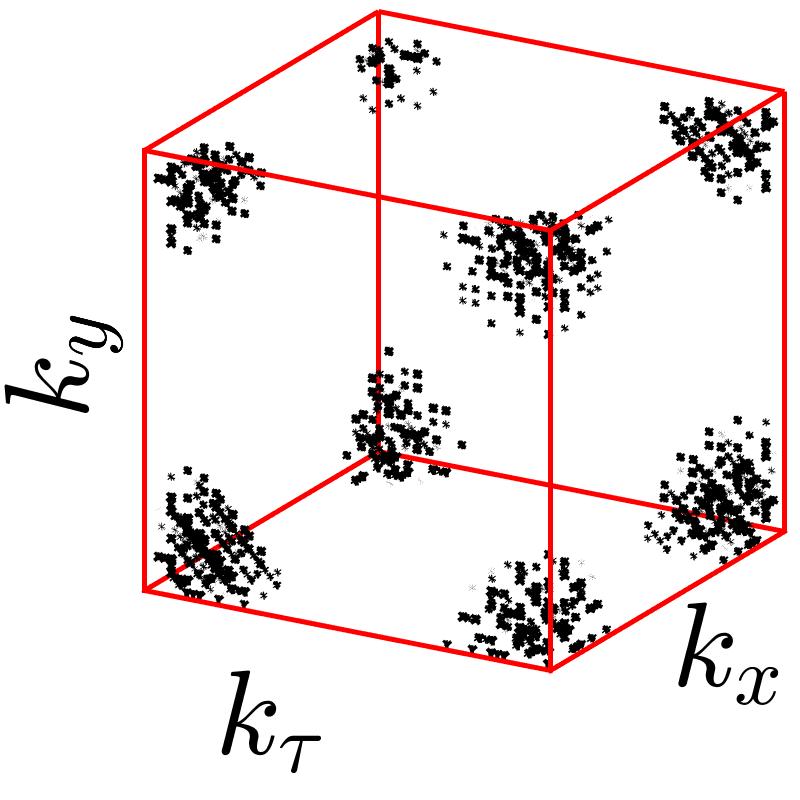}
}
\begin{picture}(0,0)
\put(-0.1,5.8){\textrm{(a)}}
\put(-0.1,2.1){\textrm{(b)}}
\put(3.05,2.1){\textrm{(c)}}
\end{picture}
\caption{(a) Bifurcation diagram of the three-dimensional periodic patterns obtained from the weakly nonlinear analysis for $\delta = 0.7$. Broken lines correspond to unstable solutions. The black dots along the bcc branch are the maximum values of the bcc solutions obtained by numerical simulations of the 3D LLE Eq. \ref{eqn:LLE}. (b) and (c) Isosurface of the intracavity field intensity corresponding to the 3D
bcc solutions of Eq. \ref{eqn:LLE} obtained from numerical simulations and its Fourier spectrum, respectively. Parameters are $E_\textrm{i}=1.05$ and $\delta = 0.7$.}
 \label{fig:Fig3}
 \end{figure}

Applying a weakly nonlinear analysis that consists of seeking nonlinear solutions by using an expansion in terms of a small parameter which measures the distance from the Turing bifurcation, it has been shown in 2D settings that only triangular or hexagonal structures are stable, and transition from hexagons to stripes is not possible for the LLE \cite{Tlidi4}. In 3D, an analytical calculation based on a weakly nonlinear analysis allows one to determine the variety and the stability properties of the three-dimensional dissipative crystals which are solutions of the generalized LLE \cite{Tlidi4,Tlidi2}. In these papers,  the solvability condition allows for the derivation of amplitude equations for the critical modes associated with a set of finite modes.  The most simple nonlinear solutions are lamellae, hexagonally
packed cylinders (hpc), and body-centered-cubic (bcc) crystals. Their
stationary solutions are $A_{lam} = (g/\alpha)^{1/2}$, 
$A_{hpc}^{\pm}= \left[-h_2 \pm [h_2^2 - 4\alpha(2h_1 - g)]^{1/2}]\right/2(2h_1 -g)$, \\
$A_{bcc}^{\pm}=\left[-h_2 \pm [h_2^2 - \alpha(2h_3 + 5h_1 - g)]^{1/2}]\right/(2h_3 + 5h_1 -g)$, with $\alpha = (E_\textrm{i} - E_{ic})/ E_{ic}(2 - \theta)^2$, $g = [2(41-30\delta)] / 9(1 - (1-\delta)^2 )^2 , h_1 = (4\delta-3)/[1 - (1-\delta)^2) 2]$,$h_2 = [1 + F(E_\textrm{i} - E_\textrm{ic})] / [E_\textrm{ic}\theta ]$, $F = [19(\delta^3-8) -4(23\delta^3 -44\delta^2-14)]/ 2(2 - \delta)^4 [1 + (2 - \delta)^2]$, $h_3 = - 2[1 +(1-\delta)^2] /[1 - (1-\theta)^2 ]^2$.  The relative linear stability analysis has been performed analytically and leads to the conclusion that only the most stable crystals are the bcc over others 3D nonlinear solutions \cite{Tlidi4}. It has been remarked in the concluding remarks of this paper that these results are obtained in a perturbative way and therefore need further support from either numerical simulations or experimental evidence'  \cite{Tlidi4}.  We fill this gap by confirming the above pattern selection scenario by numerically integrating the LLE equation with periodic boundary conditions. 

\begin{figure}
 \unitlength=70.0mm
 \centerline{
\includegraphics[width=1.50\unitlength,height=1.05\unitlength]{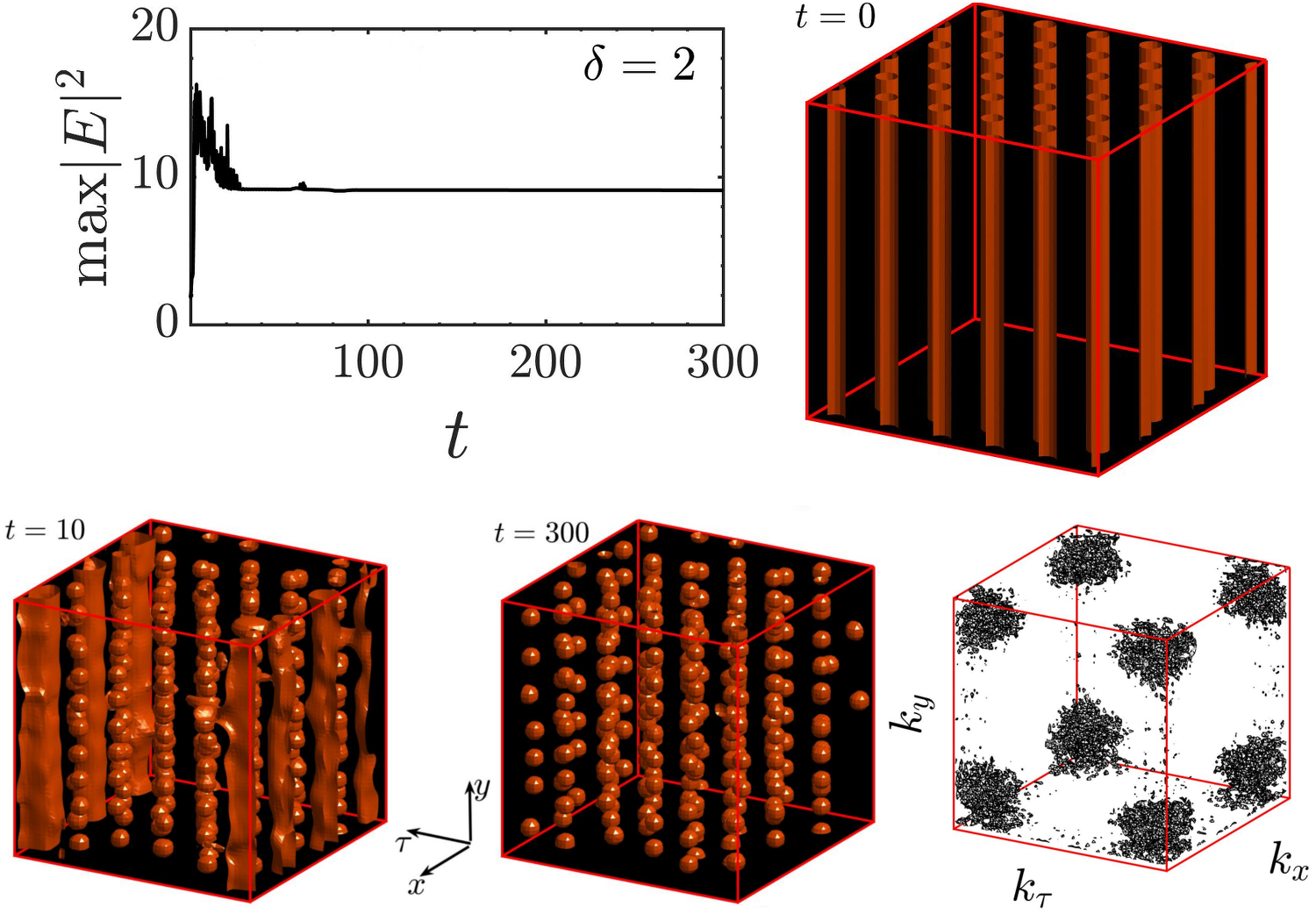}
}
\caption{Destabilization of the hexagonally packed cylinder 3D structures towards the formation bcc crystals. Parameter settings: $E_\textrm{i} = 1.42$ and $\delta = 2.0$.}
 \label{fig:Fig4}
 \end{figure}

The results of the weakly nonlinear analysis are summarized in
the bifurcation diagram displayed in Fig. \ref{fig:Fig3}(a). The stable bcc structure in Fig. \ref{fig:Fig3}(b) is obtained by time-marching the LLE using the coordinates of the associated stable wavevectors and their complex conjugates in wavenumber space, which are given by $k_c(\pm 1, \pm 1, 0)/\sqrt{2}, k_c(\pm 1, 0, \pm 1)/\sqrt{2}$ and $k_c(0, \pm 1, \pm 1)/\sqrt{2}$, as the initial condition. The Fourier transform of the bcc structures is the fcc in Fourier space as shown in Fig. \ref{fig:Fig3}(c). We can see that lamella appears supercritically. This is because the above weakly nonlinear analysis has restricted the values of
the detuning parameter to the range $\delta<\delta_{sub}$ with $\delta_{sub}<41/33$. The hpc and the bcc appear subcritically. However, lamellae and hpc are unstable, and only a branch of the bcc crystals  emerges subcritically from the homogeneous solution at the
bifurcation point. The bcc structures are unstable until a turning point given by $h_2^2 =\alpha(2h_3 + 5h_1 - g)$
is reached from which the branch $A_{bcc}$ emerges and is stable as shown in  Fig. \ref{fig:Fig3}(a). To make explicit comparisons with analytical
results, the numerical solutions are obtained for the parameter range where the system
exhibits a monostable homogeneous steady-state solution and a supercritical modulational instability. The results of numerical simulations are shown by the black dots along the bcc branch of solutions (cf. Fig. \ref{fig:Fig3}(a)). From this comparison, we can see a good agreement. To show that the bcc crystals are the most stable solutions of the 3D LLE, we choose an initial condition consisting of a hexagonally packed cylinder as shown in  Fig. \ref{fig:Fig4}. Time evolution of the system starting with this initial condition in which a  small amplitude noise is added is shown in Fig. \ref{fig:Fig4}. In an earlier stage of time evolution, the cylinders break into spheres that interact and the system reaches a stable bcc crystal.


\section{Sub-critical modulational instability and localised light bullets}\label{sec:lightbullet}

In the previous section, we have checked numerically that close to the 3D modulational instability the dynamics of the 3D Kerr cavity is predominated by the body-centered-cubic crystals over a variety of 3D structures in the cavity field intensity \cite{Tlidi2}.  The validity of this analysis is restricted to the values of the detuning parameter within the range  $\delta<\delta_{sub}$ with $\delta_{sub}<41/33$. In what follows we focus on the strongly nonlinear regime where 3D
modulational instability is subcritical, i.e., $\delta>\delta_{sub}$.  Remarkably, besides the emergence of bcc structures, the same mechanism predicts the possible existence of stable of aperiodic distribution of dissipative light bullets.  Recently, we have reported on the formation of LB in the monostable regime $\delta_{sub}<\delta<\sqrt{3}$ where the transmitted intensity, as a function of the input intensity 
is single-valued  \cite{gopalakrishnan2021dissipative}. The results reported below describe the behavior predicted based on the 3D LLE Eq. \ref{eqn:LLE}, in a strongly nonlinear regime where the HSS exhibits a bistable regime  $\delta>\sqrt{3}$.  For this purpose, we fix the detuning parameter to $\delta=2$, and we let the injected field amplitude be the control parameter. 

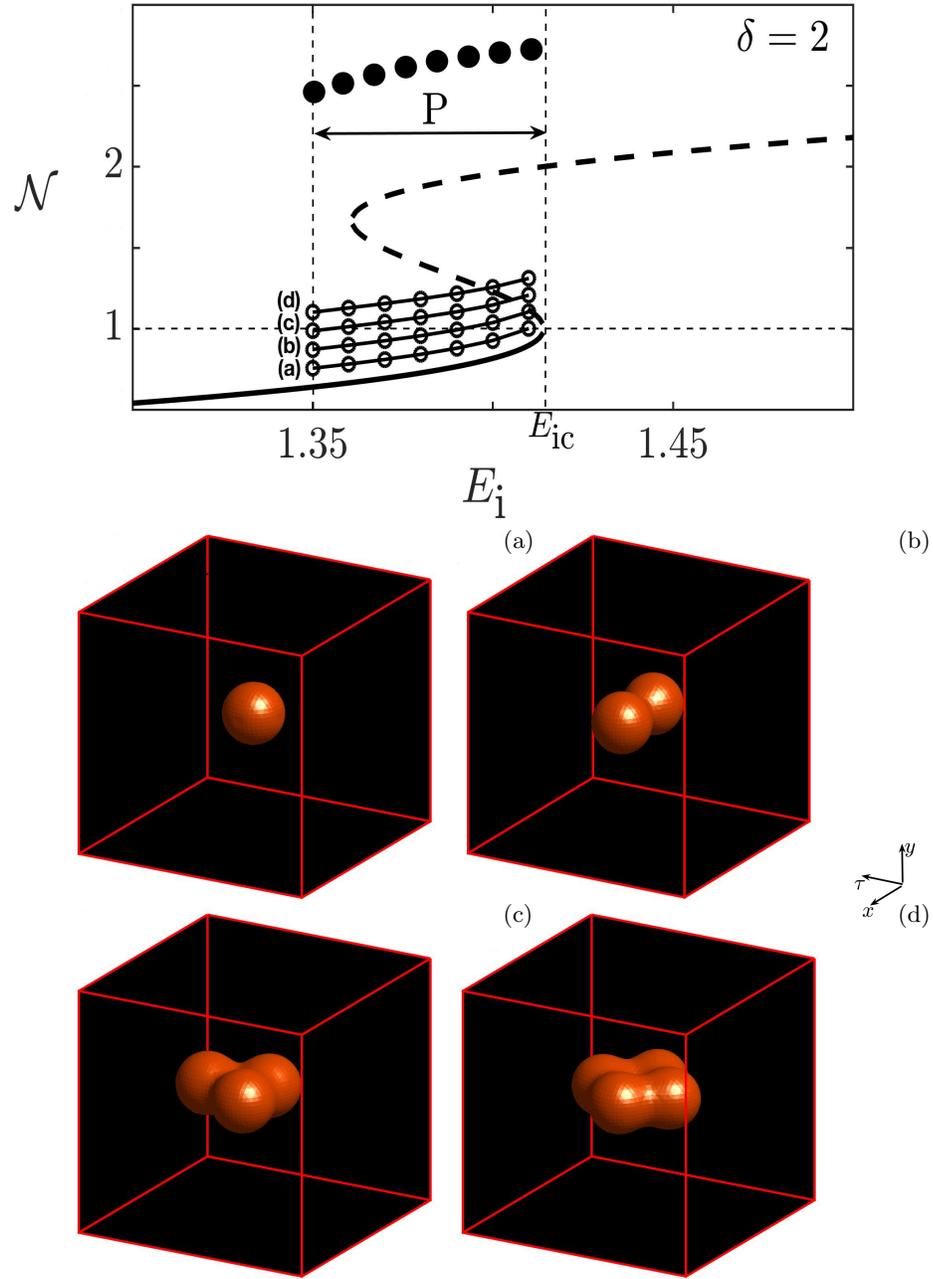
\begin{figure}
 \unitlength=60.0mm
\centerline{
\includegraphics[width=1.932\unitlength,height=1.209\unitlength]{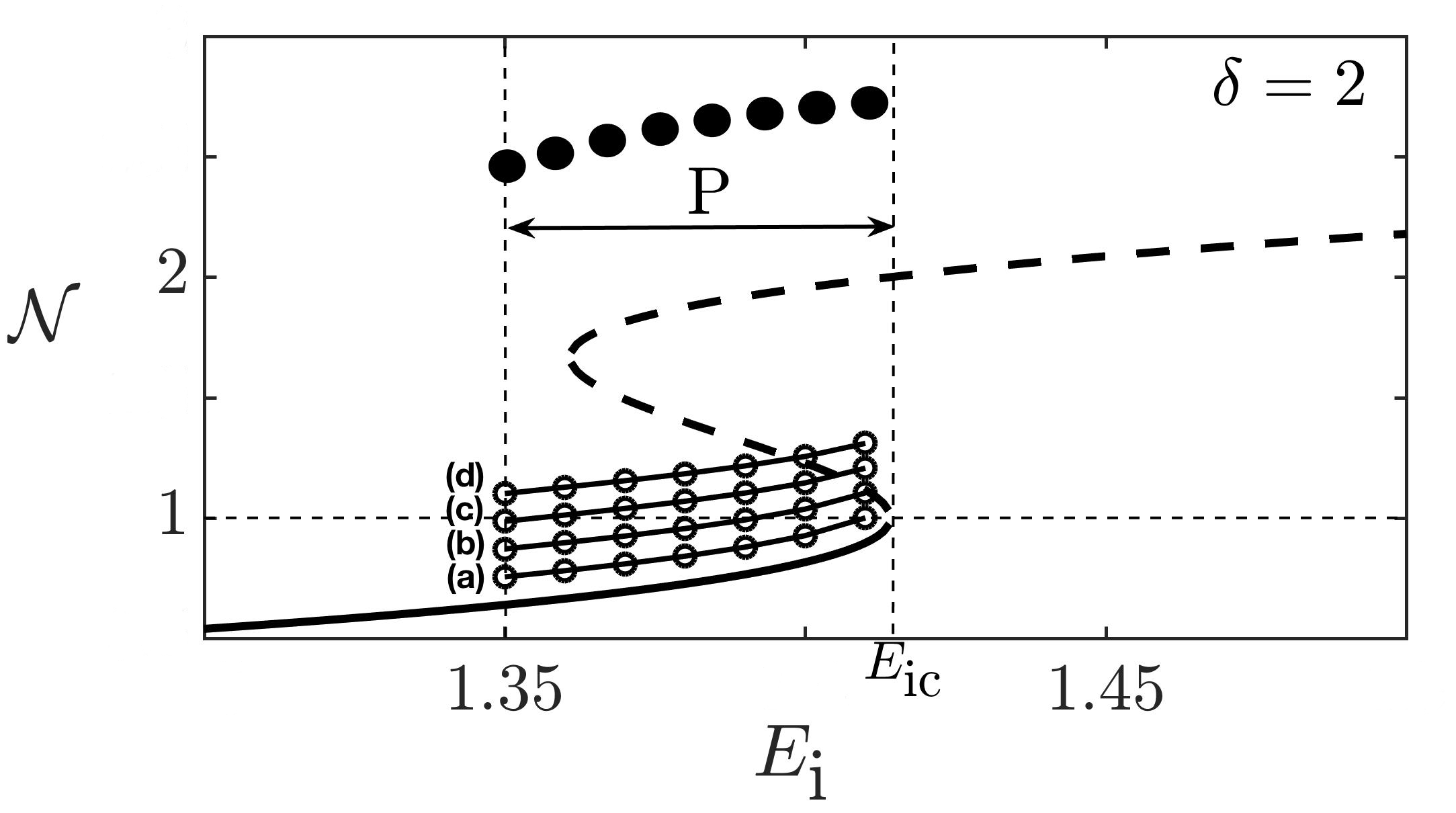}
}
 \unitlength=25.0mm
 \centerline{
\includegraphics[width=2.0\unitlength,height=2.0\unitlength]{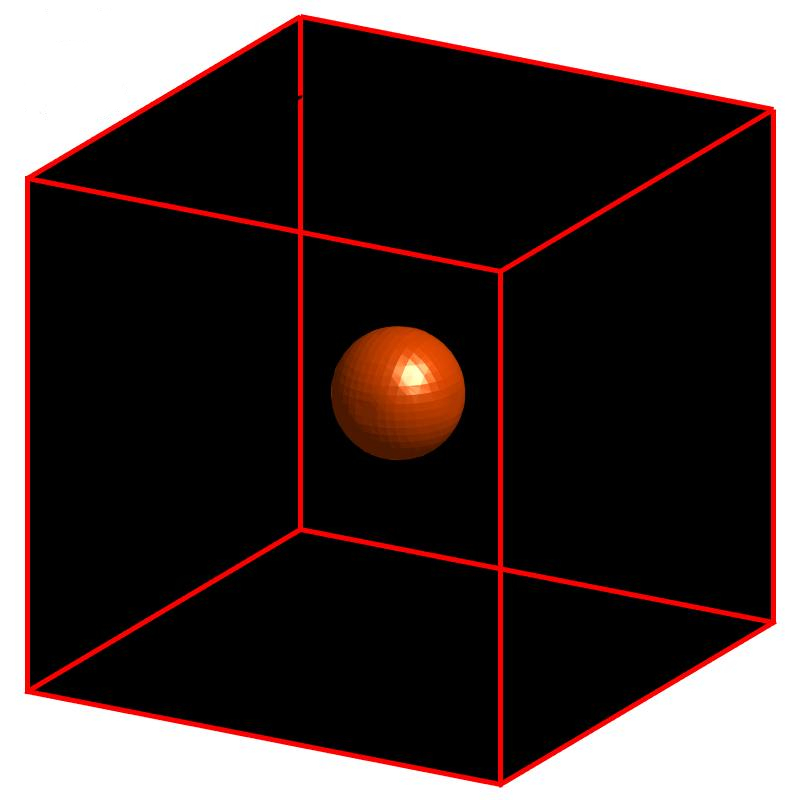}
\includegraphics[width=2.0\unitlength,height=2.0\unitlength]{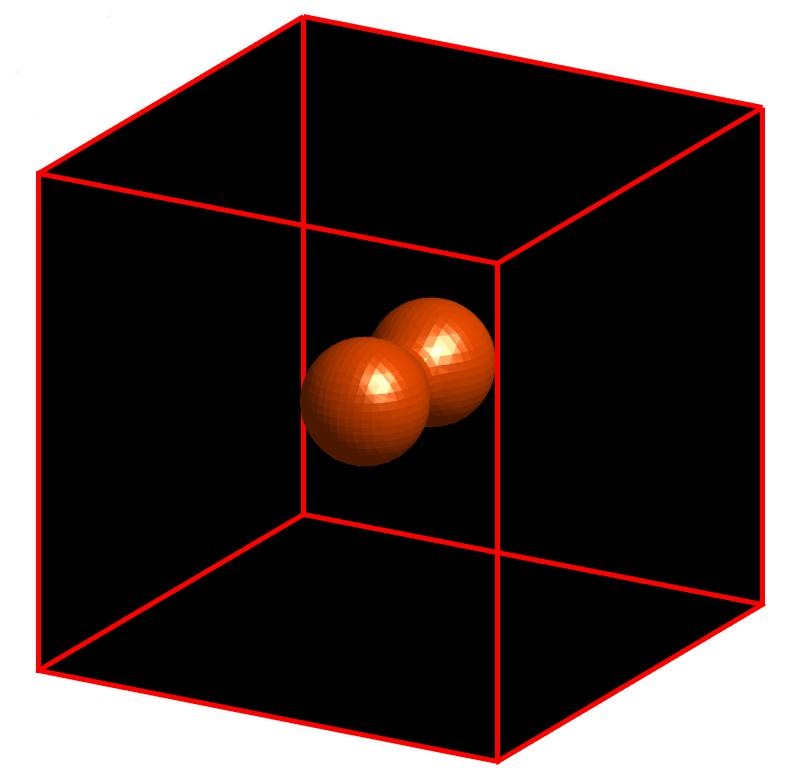}
}
 \centerline{
\includegraphics[width=2.0\unitlength,height=2.0\unitlength]{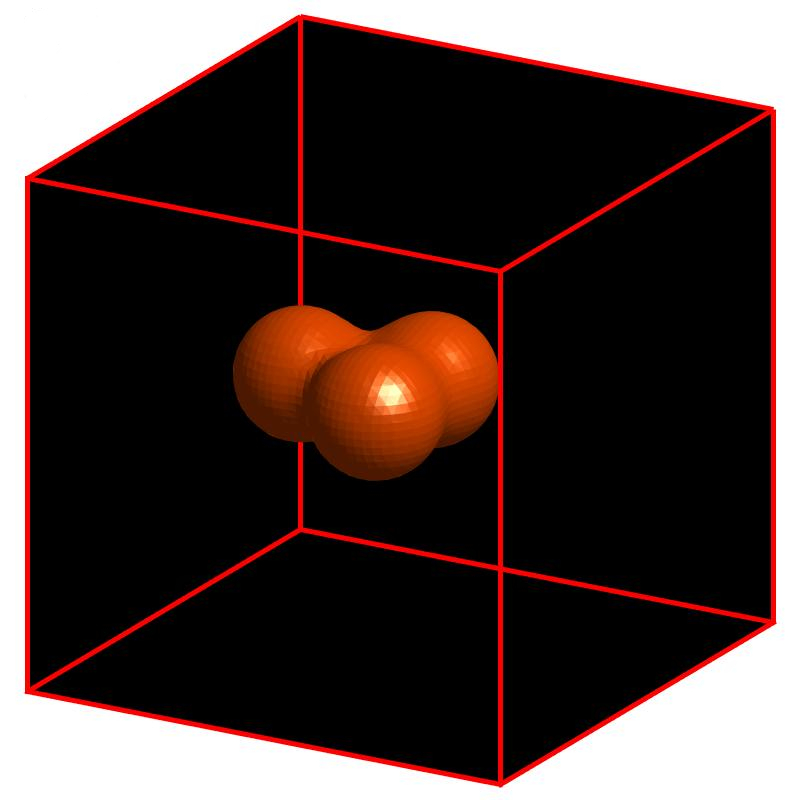}
\includegraphics[width=2.0\unitlength,height=2.0\unitlength]{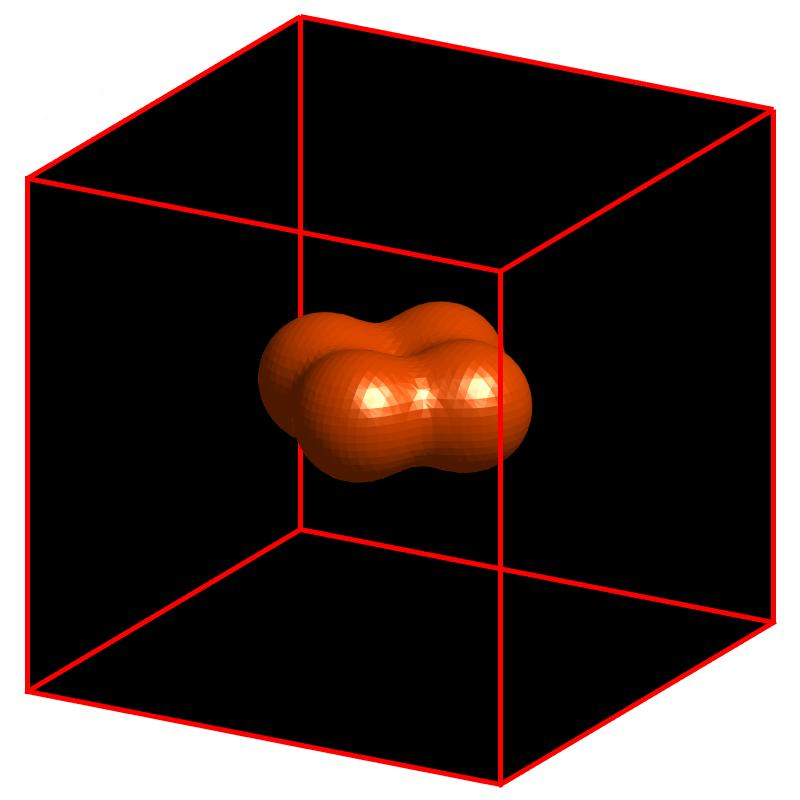}
}
 \unitlength=14.0mm
\begin{picture}(0,0)
\put(3.9,3.75){\includegraphics[width=0.6\unitlength,height=0.7\unitlength]{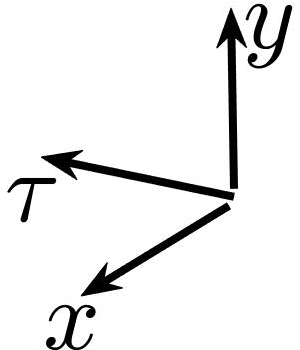}}
\end{picture}
\begin{picture}(0,0)
\put(0.5,7.25){\textrm{(a)}}
\put(4.25,7.25){\textrm{(b)}}
\put(0.5,3.7){\textrm{(c)}}
\put(4.25,3.7){\textrm{(d)}}
\end{picture}
 \caption{Bifurcation diagram associated with the LBs. The continuous black line denotes the stationary steady state. $\textrm{P}$ indicates the pinning range. The black dots represent the corresponding values for the bcc solution. (a) A single elemental LB and (b,c,d) clusters of $2$, $3$, and $4$ LBs bounded together represented using the isosurfaces of the intracavity field intensity. Parameter settings: $E_\textrm{i} = 1.40$ and $\delta = 2$.
}
 \label{fig:Fig5}
 \end{figure} 

Figure  \ref{fig:Fig5} shows the homogeneous steady states together with the extended and periodic 3D dissipative structures. Numerical simulations showed that in the strongly nonlinear regime the bcc structures are the most stable crystals.  The $L_2$ norm, defined below by Eq. \ref{L2norm}, is plotted as a function of the injected field amplitude in Fig. \ref{fig:Fig5}.  In this bifurcation diagram, the homogeneous steady states are plotted together with the bcc crystals. The upper branch of the bistable HSS curve is entirely unstable for the 3D modulation instability denoted by the dashed black line in Fig. \ref{fig:Fig5}.  The $L_2$ norm associated with the bcc crystals is indicated by the black dot in Fig. \ref{fig:Fig5}. From this figure, we see a domain indicated by $P$ where the system exhibits multistability. The lower homogeneous steady states are represented by a continuous black line, and along with the bcc crystal, additional variety of aperiodic 3D structures can be obtained.
Figure \ref{fig:Fig5}(a) shows a single stationary LB obtained numerically by using an initial condition consisting of a Gaussian shell  centered in the computational domain.
By placing multiple Gaussian shells and by varying the distances between them, one can obtain multiple robust stationary LBs within the optical cavity. Figure \ref{fig:Fig5} shows two, three and four light bullets bounded together along different perspectives for $E_\textrm{i} = 1.40$, and $\delta = 2$. Indeed, once robust LBs have been obtained for a specific parameter setting, numerically they are used as the initial condition for further simulations, for instance, to obtain the bifurcation diagram for varying values of the amplitude of the injected field $E_\textrm{i}$. Since the
amplitudes of LBs having different numbers of 3D peaks are more or less the same, it is convenient to plot the dimensionless ``$L_2$ norm'',
\begin{equation}
{\cal{N}}= \int\vert E - E_s \vert^2 dx\,dy\,d\tau
\label{L2norm}
\end{equation}
as a function of the injected field amplitude $E_\textrm{i}$.  Curves a,b,c,d in Fig. \ref{fig:Fig5}(a) shows the $L_2$ norm associated with LBs with $1$, $2$, $3$, and $4$ peaks.   The single LB obtained by a direct numerical simulation of Eq. (\ref{eqn:LLE}),  can be solved under a spherical approximation. This approximation appears plausible since the LB is a stationary object with a spherical symmetry, and has the form $E(r) = E_s(1 + A(r))$ with $r=(x^2+y^2+\tau^2)^{1/2}$ and $E_s$ denotes the lower homogeneous steady state.  By replacing this ansatz in the 3D LLE Eq. \ref{eqn:LLE}, and decomposing the intracavity field  into real and imaginary parts as $E_S(r,z) = A_r + \dot{\imath} A_i(r)$, and by replacing the Laplace operator in polar coordinates
 $\nabla_{\perp}^2  + \partial^2/\tau^2=\partial^2/\partial r^2 + (2/r) \partial/\partial r$, we obtain four first order ODEs as
\begin{eqnarray}
\frac{d y_1}{dr} &=& y_2, \nonumber \\
\frac{d y_2}{dr} &=& \delta y_1 + y_3 - \Big(y^{2}_{1} + y^{2}_{3}\Big) y_1 - \frac{2 y_2}{r}, \nonumber \\
\frac{d y_3}{dr} &=& y_4, \nonumber \\
\frac{d y_4}{dr} &=& E_\textrm{i} - y_1 + \delta y_3 - \Big(y^{2}_{1} + y^{2}_{3}\Big) y_3 - \frac{2 y_4}{r}.
\label{bvpLL}
\end{eqnarray}
where $y_1 = A_r, y_2 = \frac{d A_r}{dr}, y_3 = A_i, y_4 = \frac{d A_i}{dr}$. We solve these set of equations as a boundary value problem  for a finite value of $A_r$ at $r = 0$, and $A_r, A_i$ $\rightarrow 0$ as $r \rightarrow \infty$.

  \begin{figure}
 \unitlength=14.0mm
\centerline{
\includegraphics[width=3.0\unitlength,height=2.0\unitlength]{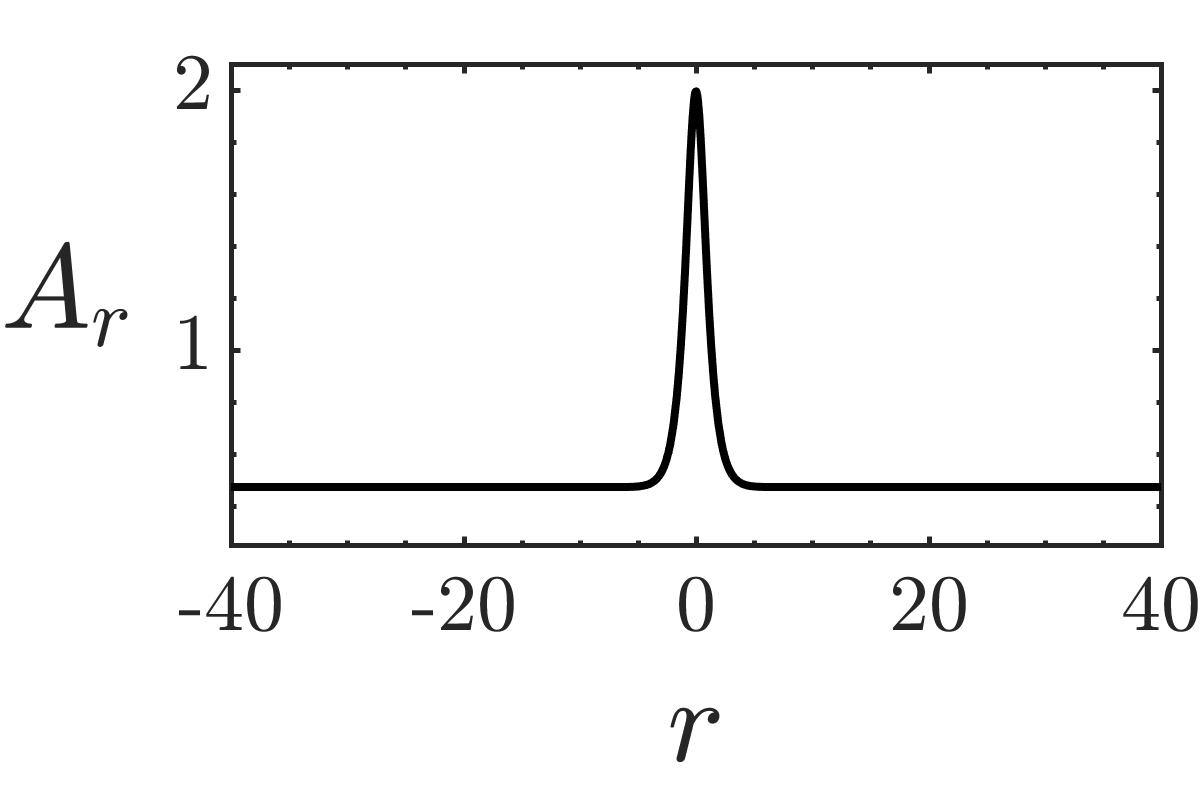}
\includegraphics[width=3.0\unitlength,height=2.0\unitlength]{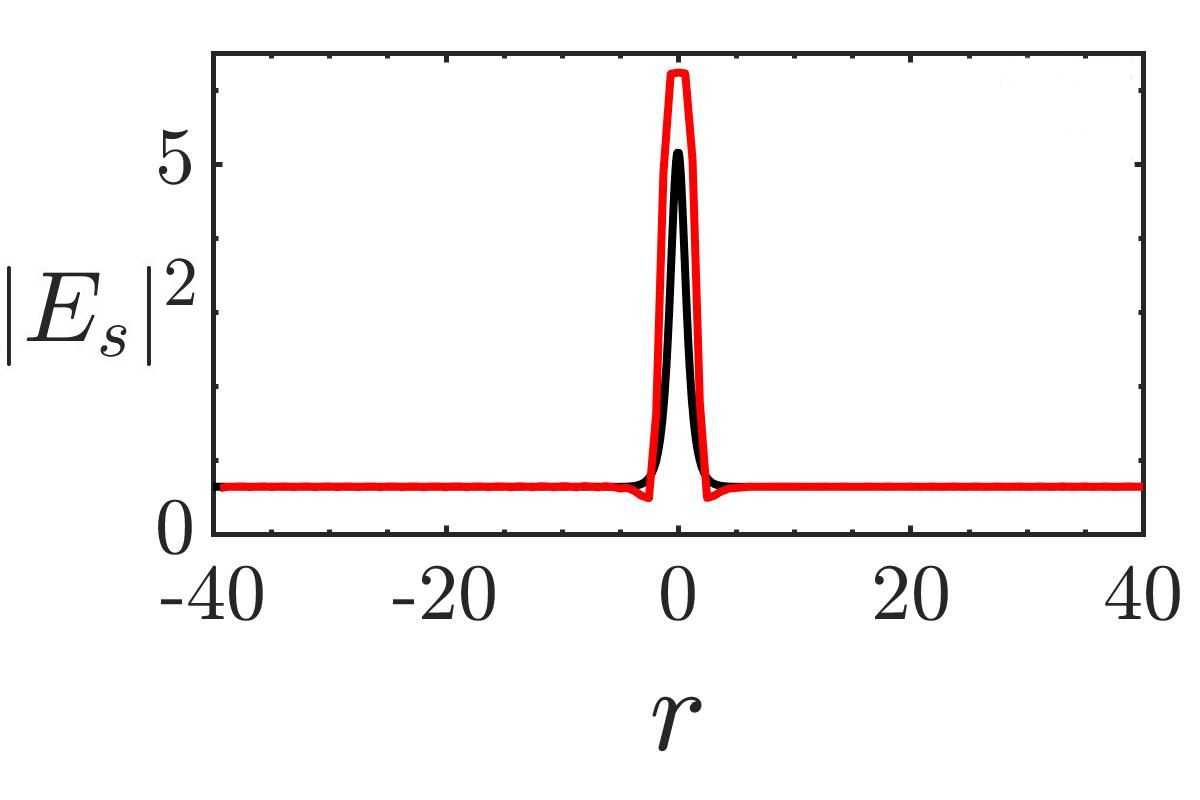}
}
\centerline{
\includegraphics[width=3.0\unitlength,height=2.0\unitlength]{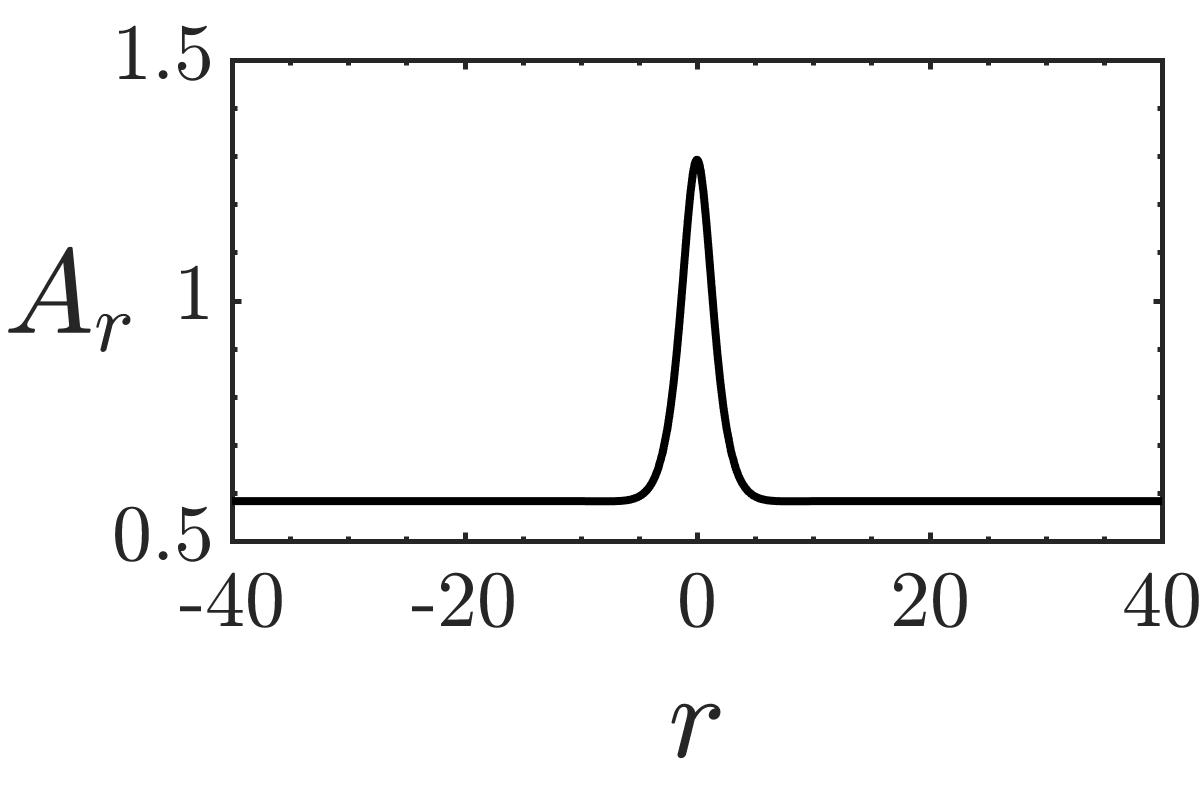}
\includegraphics[width=3.0\unitlength,height=2.0\unitlength]{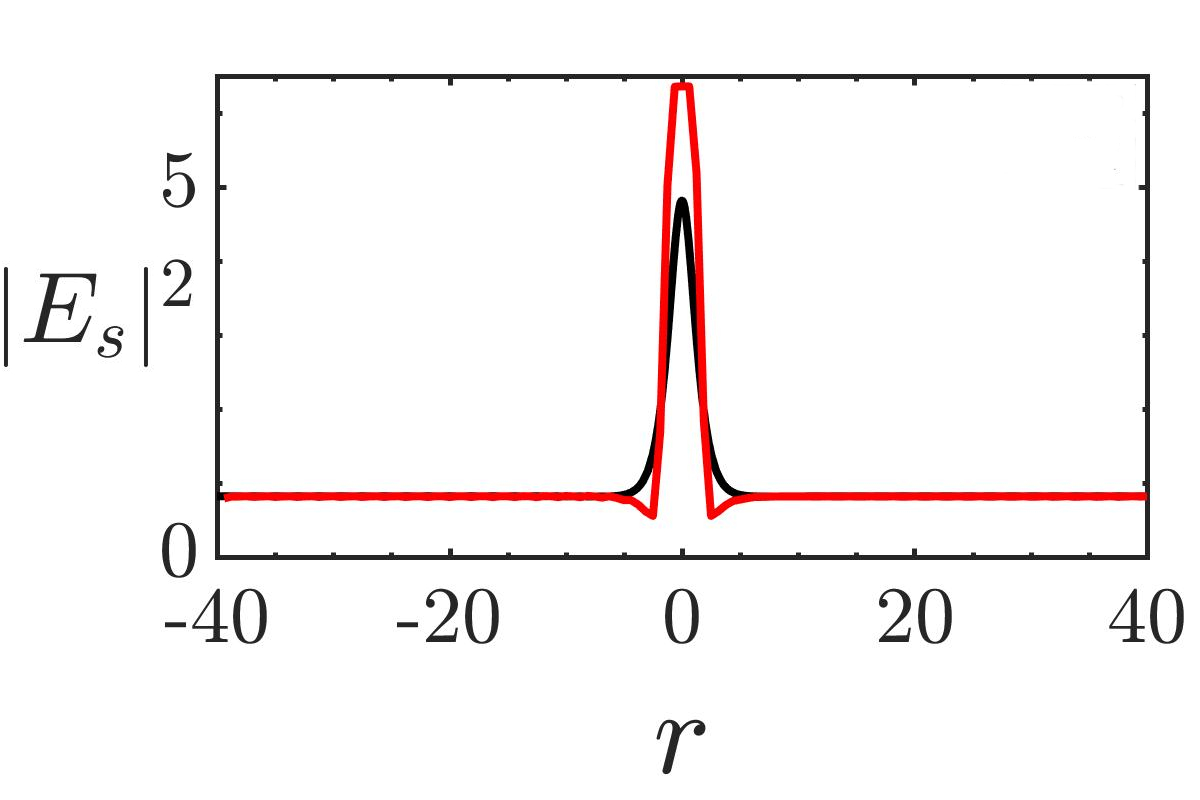}
}
\begin{picture}(0,0)
\put(0.9,4.3){\textrm{(a)}}
\put(4.25,4.3){\textrm{(b)}}
\put(0.9,2.2){\textrm{(c)}}
\put(4.25,2.2){\textrm{(d)}}
\end{picture}
 \caption{The real part of the steady-state solutions obtained using spherical symmetry considerations is shown in panels (a) and (c), with the total intensity shown using a black line in panels (b) and (d). The steady-state solution obtained by time-marching the LLE as discussed in the Letter is shown using a red line in panels (b) and (d) for comparison. Parameter settings are $\delta = 2$ (a) $E_\textrm{i} = 1.35$ (b) $E_\textrm{i}= 1.40$.
}
 \label{fig:Fig6}
 \end{figure}
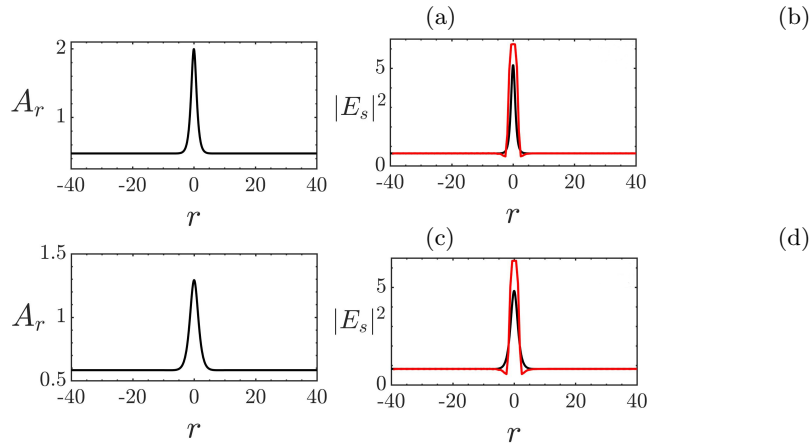 
 
Figure \ref{fig:Fig6} shows the steady-state solution obtained by solving the above set of equations with the real part $A_r$ shown in panels (a, b), and $\vert E_S \vert^2$ in (b, d) in black respectively for $\delta = 2.0$, and for two values of $E_\textrm{i}$. The solution obtained via integrating the LLE, is shown in red in panels (b, d) for comparison. The steady-state solutions obtained as a boundary value problem involves a careful choice in the initial amplitude for $A_r$ at $r = 0$, and by imposing smooth conditions at $r = 0$ as $dA_r/dr = dA_i/dr = 0$. The present method has been used in earlier studies, for instance in \cite{Edmundson1} where the study focussed on the nonlinear Schr\"{o}dinger equation. It can be noted in panels (b, d) that the oscillatory tail is more evident in the solutions obtained numerically by time-marching the LLE.  although they are evident in both the solutions. Though the solutions are in good qualitative agreement, the absolute values of the intensity obtained using the two methods are slightly different. The steady-state solution obtained using spherical symmetry considerations by solving equations \ref{bvpLL} as a boundary value problem has been carried out to qualitatively validate the results from the nonlinear simulations, rather than for precise quantitative comparisons.

 \begin{figure}
 \unitlength=50.0mm
 \centerline{
\includegraphics[width=1.0\unitlength,height=1.0\unitlength]{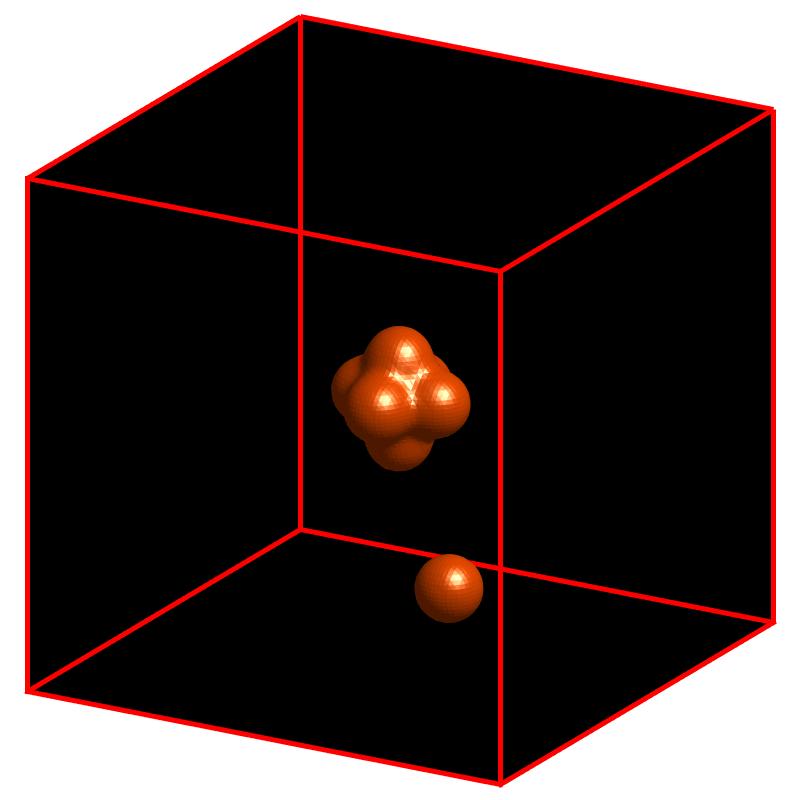}
}
\caption{Clusters of LBs coexisting along with an isolated LB. Parameter settings: $E_\textrm{i} = 1.40$, $\delta = 2$.}
 \label{fig:Fig5bis}
 \end{figure}

It should be noted that the LBs can bind themselves to each other via their oscillatory tails.  We focus now on the situation where 3D peaks are close-packed so that their overlapping oscillatory tails interact strongly. All
these LBs coexist as stable solutions with the bcc crystals in
the range $\textrm{P}$ shown in Fig. \ref{fig:Fig5}(a). The LBs are localized dissipative structures along the $x$, $y$, and $\tau$ directions. They can be seen as
a cluster of the elemental structure (a single LB) with a well-defined size.  Their position depends on the initial conditions, 
and the maximum of the coexisting LBs is essentially constant for fixed values of the system parameters.

As a  final example of what 3D LLE Eq. \ref{eqn:LLE} is able to generate, is clusters of LBs can coexist with a single isolated LB as shown in Fig. \ref{fig:Fig5bis}. Depending on the initial condition, we have also been
able to find stable LBs with a few  3D peaks packed together forming a cluster of six LBs bounded together coexisting with a single LB.  [see Fig. \ref{fig:Fig5bis}]. These two 3D localized objects are far away from each other.  The distance between them is determined solely by the initial conditions used.






\section{Discussion and concluding remarks}\label{sec:discconc}
In conclusion, we have shown that the 3D lugiato-Lefever equation  Eq. \ref{eqn:LLE}, captures quite a  large variety of three-dimensional dissipative structures and clusters of light-bullets.  In the first part, we have discussed the 3D pattern selection through the weakly nonlinear analysis and the relative stability analysis. We have checked numerically that the only possible periodic structures  are the body-centered-cubic crystals. This n-analysis is restricted to the weakly nonlinear regime where close to the 3D modulational instability appears super-criticaly. The  second part has been centered rather on the formation of light-bullet and clusters of them. We have shown that there exists a range of parameters called pinning zone where the system exhibits a multistability behavior. Besides the body-centered cubic crystals and the homogeneous steady states, another type of localized and aperiodic 3D structures have been generated for a fixed value of the system's parameters.

The multiplicity of
these 3D solutions of the LLE is strongly reminiscent of homoclinic snaking. The full diagram can be complex, and we displayed only four
branches of LBs. Indeed, in one-dimensional settings, localized structures exhibit a homoclinic snaking bifurcation which has been established by continuation algorithms \cite{gomila2007bifurcation,tlidi2010high}. The snaking bifurcation diagram consists of two snaking curves: one describes localized structures with $2n$ peaks, while the other corresponds to $2n + 1$ peaks where $n$ is a positive integer. As one moves further along the snaking curve, the LB becomes better localized and acquires stability at the turning point where the slope becomes infinite. Outside of the pinning range, the LB begins to grow by adding extra peaks symmetrically at either side. This growth is associated with back and forth oscillations across the pinning range of the control parameter. This homoclinic snaking has been established first in  \cite{woods_heteroclinic_1999}. An extension to two-dimensional settings of the homoclinic bifurcation has been discussed in recent overviews \cite{lloyd2008localized,knobloch2015spatial}. However, continuation algorithms in 3D are still largely unexplored, and most of the results are obtained by direct numerical simulations of the governing equation.

\section*{Funding}

 K.P. acknowledges the support by the Fonds Wetenschappelijk Onderzoek-Vlaanderen FWO (G0E5819N) and the Methusalem
Foundation.  We also acknowledge the support from the French National Research Agency (LABEX CEMPI, Grant No. ANR-11- LABX-0007) as well as the French Ministry of Higher Education and Research, Hauts de France council and European Regional Development Fund (ERDF) through the Contrat de Projets Etat-Region (CPER Photonics for Society P4S). M.T acknowledges financial support from the Fonds de la Recherche Scientifique FNRS under Grant CDR no. 35333527 "Semiconductor optical comb generator". A part of this work was supported by the "Laboratoire Associ\'{e} International" University of Lille - ULB on "Self-organisation of light and extreme events" (LAI-ALLURE)."

\bibliography{BiblioLLE3D}

\providecommand{\noopsort}[1]{}\providecommand{\singleletter}[1]{#1}%
\begin{thebibliography}{57}%
\makeatletter
\providecommand \@ifxundefined [1]{%
 \@ifx{#1\undefined}
}%
\providecommand \@ifnum [1]{%
 \ifnum #1\expandafter \@firstoftwo
 \else \expandafter \@secondoftwo
 \fi
}%
\providecommand \@ifx [1]{%
 \ifx #1\expandafter \@firstoftwo
 \else \expandafter \@secondoftwo
 \fi
}%
\providecommand \natexlab [1]{#1}%
\providecommand \enquote  [1]{``#1''}%
\providecommand \bibnamefont  [1]{#1}%
\providecommand \bibfnamefont [1]{#1}%
\providecommand \citenamefont [1]{#1}%
\providecommand \href@noop [0]{\@secondoftwo}%
\providecommand \href [0]{\begingroup \@sanitize@url \@href}%
\providecommand \@href[1]{\@@startlink{#1}\@@href}%
\providecommand \@@href[1]{\endgroup#1\@@endlink}%
\providecommand \@sanitize@url [0]{\catcode `\\12\catcode `\$12\catcode
  `\&12\catcode `\#12\catcode `\^12\catcode `\_12\catcode `\%12\relax}%
\providecommand \@@startlink[1]{}%
\providecommand \@@endlink[0]{}%
\providecommand \url  [0]{\begingroup\@sanitize@url \@url }%
\providecommand \@url [1]{\endgroup\@href {#1}{\urlprefix }}%
\providecommand \urlprefix  [0]{URL }%
\providecommand \Eprint [0]{\href }%
\providecommand \doibase [0]{http://dx.doi.org/}%
\providecommand \selectlanguage [0]{\@gobble}%
\providecommand \bibinfo  [0]{\@secondoftwo}%
\providecommand \bibfield  [0]{\@secondoftwo}%
\providecommand \translation [1]{[#1]}%
\providecommand \BibitemOpen [0]{}%
\providecommand \bibitemStop [0]{}%
\providecommand \bibitemNoStop [0]{.\EOS\space}%
\providecommand \EOS [0]{\spacefactor3000\relax}%
\providecommand \BibitemShut  [1]{\csname bibitem#1\endcsname}%
\let\auto@bib@innerbib\@empty
\bibitem [{\citenamefont {Cross}\ and\ \citenamefont
  {Hohenberg}(1993)}]{cross1993pattern}%
  \BibitemOpen
  \bibfield  {author} {\bibinfo {author} {\bibfnamefont {Mark~C}\ \bibnamefont
  {Cross}}\ and\ \bibinfo {author} {\bibfnamefont {Pierre~C}\ \bibnamefont
  {Hohenberg}},\ }\bibfield  {title} {\enquote {\bibinfo {title} {Pattern
  formation outside of equilibrium},}\ }\href@noop {} {\bibfield  {journal}
  {\bibinfo  {journal} {Reviews of modern physics}\ }\textbf {\bibinfo {volume}
  {65}},\ \bibinfo {pages} {851} (\bibinfo {year} {1993})}\BibitemShut
  {NoStop}%
\bibitem [{\citenamefont {Arecchi}\ \emph {et~al.}(1999)\citenamefont
  {Arecchi}, \citenamefont {Boccaletti},\ and\ \citenamefont
  {Ramazza}}]{arecchi1999pattern}%
  \BibitemOpen
  \bibfield  {author} {\bibinfo {author} {\bibfnamefont {F~Tito}\ \bibnamefont
  {Arecchi}}, \bibinfo {author} {\bibfnamefont {Stefano}\ \bibnamefont
  {Boccaletti}}, \ and\ \bibinfo {author} {\bibfnamefont {PierLuigi}\
  \bibnamefont {Ramazza}},\ }\bibfield  {title} {\enquote {\bibinfo {title}
  {Pattern formation and competition in nonlinear optics},}\ }\href@noop {}
  {\bibfield  {journal} {\bibinfo  {journal} {Physics Reports}\ }\textbf
  {\bibinfo {volume} {318}},\ \bibinfo {pages} {1--83} (\bibinfo {year}
  {1999})}\BibitemShut {NoStop}%
\bibitem [{\citenamefont {Staliunas}\ and\ \citenamefont
  {Sánchez-Morcillo}(2003)}]{Staliunas03}%
  \BibitemOpen
  \bibfield  {author} {\bibinfo {author} {\bibfnamefont {K.}~\bibnamefont
  {Staliunas}}\ and\ \bibinfo {author} {\bibfnamefont {V.}~\bibnamefont
  {Sánchez-Morcillo}},\ }\href@noop {} {\emph {\bibinfo {title} {Transverse
  patterns in nonlinear optical resonators. {S}pringer {T}racts in {M}odern
  {P}hysics}}}\ (\bibinfo  {publisher} {{B}erlin, {G}ermany: {S}pringer.},\
  \bibinfo {year} {2003})\BibitemShut {NoStop}%
\bibitem [{\citenamefont {Murray}(2007)}]{murray2007mathematical}%
  \BibitemOpen
  \bibfield  {author} {\bibinfo {author} {\bibfnamefont {James~D}\ \bibnamefont
  {Murray}},\ }\href@noop {} {\emph {\bibinfo {title} {Mathematical biology: I.
  An introduction}}},\ Vol.~\bibinfo {volume} {17}\ (\bibinfo  {publisher}
  {Springer Science \& Business Media},\ \bibinfo {year} {2007})\BibitemShut
  {NoStop}%
\bibitem [{\citenamefont {Akhmediev}\ and\ \citenamefont
  {(eds).}(2008)}]{Akhmediev08}%
  \BibitemOpen
  \bibfield  {author} {\bibinfo {author} {\bibfnamefont {N.}~\bibnamefont
  {Akhmediev}}\ and\ \bibinfo {author} {\bibfnamefont {A.~Ankiewicz}\
  \bibnamefont {(eds).}},\ }\href@noop {} {\emph {\bibinfo {title} {Dissipative
  solitons: from optics to biology and medicine.}}}\ (\bibinfo  {publisher}
  {Lecture Notes in Physics, vol. 751. Heidelberg, Germany: Springer.},\
  \bibinfo {year} {2008})\BibitemShut {NoStop}%
\bibitem [{\citenamefont {Tlidi}\ \emph {et~al.}(2014)\citenamefont {Tlidi},
  \citenamefont {Staliunas}, \citenamefont {Panajotov}, \citenamefont
  {Vladimirov},\ and\ \citenamefont {Clerc}}]{Tlidi5}%
  \BibitemOpen
  \bibfield  {author} {\bibinfo {author} {\bibfnamefont {M.}~\bibnamefont
  {Tlidi}}, \bibinfo {author} {\bibfnamefont {K.}~\bibnamefont {Staliunas}},
  \bibinfo {author} {\bibfnamefont {K.}~\bibnamefont {Panajotov}}, \bibinfo
  {author} {\bibfnamefont {A.~G.}\ \bibnamefont {Vladimirov}}, \ and\ \bibinfo
  {author} {\bibfnamefont {M.~G.}\ \bibnamefont {Clerc}},\ }\bibfield  {title}
  {\enquote {\bibinfo {title} {Localized structures in dissipative media: from
  optics to plant ecology},}\ }\href@noop {} {\bibfield  {journal} {\bibinfo
  {journal} {Phil. trans. R. Soc. A}\ }\textbf {\bibinfo {volume} {372}},\
  \bibinfo {pages} {20140101} (\bibinfo {year} {2014})}\BibitemShut {NoStop}%
\bibitem [{\citenamefont {Tlidi}\ and\ \citenamefont
  {Clerc}(2016)}]{tlidi2016nonlinear}%
  \BibitemOpen
  \bibfield  {author} {\bibinfo {author} {\bibfnamefont {Mustapha}\
  \bibnamefont {Tlidi}}\ and\ \bibinfo {author} {\bibfnamefont {Marcel~G}\
  \bibnamefont {Clerc}},\ }\bibfield  {title} {\enquote {\bibinfo {title}
  {Nonlinear dynamics: Materials, theory and experiments},}\ }\href@noop {}
  {\bibfield  {journal} {\bibinfo  {journal} {Springer Proceedings in Physics}\
  }\textbf {\bibinfo {volume} {173}} (\bibinfo {year} {2016})}\BibitemShut
  {NoStop}%
\bibitem [{\citenamefont {Lugiato}\ \emph {et~al.}(2018)\citenamefont
  {Lugiato}, \citenamefont {Prati}, \citenamefont {Gorodetsky},\ and\
  \citenamefont {Kippenberg}}]{lugiato2018kenberberg}%
  \BibitemOpen
  \bibfield  {author} {\bibinfo {author} {\bibfnamefont {LA}~\bibnamefont
  {Lugiato}}, \bibinfo {author} {\bibfnamefont {F}~\bibnamefont {Prati}},
  \bibinfo {author} {\bibfnamefont {ML}~\bibnamefont {Gorodetsky}}, \ and\
  \bibinfo {author} {\bibfnamefont {TJ}~\bibnamefont {Kippenberg}},\ }\bibfield
   {title} {\enquote {\bibinfo {title} {From the {L}ugiato--{L}efever equation
  to microresonator-based soliton kerr frequency combs},}\ }\href@noop {}
  {\bibfield  {journal} {\bibinfo  {journal} {Philosophical Transactions of the
  Royal Society A: Mathematical, Physical and Engineering Sciences}\ }\textbf
  {\bibinfo {volume} {376}},\ \bibinfo {pages} {20180113} (\bibinfo {year}
  {2018})}\BibitemShut {NoStop}%
\bibitem [{\citenamefont {Tlidi}\ \emph {et~al.}(2018)\citenamefont {Tlidi},
  \citenamefont {Clerc},\ and\ \citenamefont
  {Panajotov}}]{tlidi2018dissipative}%
  \BibitemOpen
  \bibfield  {author} {\bibinfo {author} {\bibfnamefont {Mustapha}\
  \bibnamefont {Tlidi}}, \bibinfo {author} {\bibfnamefont {MG}~\bibnamefont
  {Clerc}}, \ and\ \bibinfo {author} {\bibfnamefont {Krassimir}\ \bibnamefont
  {Panajotov}},\ }\href@noop {} {\enquote {\bibinfo {title} {Dissipative
  structures in matter out of equilibrium: from chemistry, photonics and
  biology, the legacy of ilya prigogine (part 1)},}\ } (\bibinfo {year}
  {2018})\BibitemShut {NoStop}%
\bibitem [{\citenamefont {Fortier}\ and\ \citenamefont
  {Baumann}(2019)}]{fortier201920}%
  \BibitemOpen
  \bibfield  {author} {\bibinfo {author} {\bibfnamefont {Tara}\ \bibnamefont
  {Fortier}}\ and\ \bibinfo {author} {\bibfnamefont {Esther}\ \bibnamefont
  {Baumann}},\ }\bibfield  {title} {\enquote {\bibinfo {title} {20 years of
  developments in optical frequency comb technology and applications},}\
  }\href@noop {} {\bibfield  {journal} {\bibinfo  {journal} {Communications
  Physics}\ }\textbf {\bibinfo {volume} {2}},\ \bibinfo {pages} {1--16}
  (\bibinfo {year} {2019})}\BibitemShut {NoStop}%
\bibitem [{\citenamefont {Scorggie}\ \emph {et~al.}(1994)\citenamefont
  {Scorggie}, \citenamefont {Firth}, \citenamefont {Mc{D}onald}, \citenamefont
  {Tlidi}, \citenamefont {Lefever},\ and\ \citenamefont {Lugiato}}]{Scroggie1}%
  \BibitemOpen
  \bibfield  {author} {\bibinfo {author} {\bibfnamefont {A.~J.}\ \bibnamefont
  {Scorggie}}, \bibinfo {author} {\bibfnamefont {W.~J.}\ \bibnamefont {Firth}},
  \bibinfo {author} {\bibfnamefont {G.~S.}\ \bibnamefont {Mc{D}onald}},
  \bibinfo {author} {\bibfnamefont {M.}~\bibnamefont {Tlidi}}, \bibinfo
  {author} {\bibfnamefont {R.}~\bibnamefont {Lefever}}, \ and\ \bibinfo
  {author} {\bibfnamefont {L.~A.}\ \bibnamefont {Lugiato}},\ }\bibfield
  {title} {\enquote {\bibinfo {title} {Pattern formation in a passive {K}err
  cavity},}\ }\href@noop {} {\bibfield  {journal} {\bibinfo  {journal} {Chaos
  Solitons Fract.}\ }\textbf {\bibinfo {volume} {4}},\ \bibinfo {pages} {1323}
  (\bibinfo {year} {1994})}\BibitemShut {NoStop}%
\bibitem [{\citenamefont {Taranenko}\ \emph {et~al.}(2000)\citenamefont
  {Taranenko}, \citenamefont {Ganne}, \citenamefont {Kuszelewicz},\ and\
  \citenamefont {Weiss}}]{Taranenko_pra00}%
  \BibitemOpen
  \bibfield  {author} {\bibinfo {author} {\bibfnamefont {V.~B.}\ \bibnamefont
  {Taranenko}}, \bibinfo {author} {\bibfnamefont {I.}~\bibnamefont {Ganne}},
  \bibinfo {author} {\bibfnamefont {R.~J.}\ \bibnamefont {Kuszelewicz}}, \ and\
  \bibinfo {author} {\bibfnamefont {C.~O.}\ \bibnamefont {Weiss}},\ }\bibfield
  {title} {\enquote {\bibinfo {title} {Patterns and localized structures in
  bistable semiconductor resonators},}\ }\href {\doibase
  10.1103/PhysRevA.61.063818} {\bibfield  {journal} {\bibinfo  {journal} {Phys.
  Rev. A}\ }\textbf {\bibinfo {volume} {61}},\ \bibinfo {pages} {063818}
  (\bibinfo {year} {2000})}\BibitemShut {NoStop}%
\bibitem [{\citenamefont {Taranenko}\ \emph {et~al.}(2001)\citenamefont
  {Taranenko}, \citenamefont {Ganne}, \citenamefont {Kuszelewicz},\ and\
  \citenamefont {Weiss}}]{taranenko2001spatial}%
  \BibitemOpen
  \bibfield  {author} {\bibinfo {author} {\bibfnamefont {VB}~\bibnamefont
  {Taranenko}}, \bibinfo {author} {\bibfnamefont {I}~\bibnamefont {Ganne}},
  \bibinfo {author} {\bibfnamefont {R}~\bibnamefont {Kuszelewicz}}, \ and\
  \bibinfo {author} {\bibfnamefont {CO}~\bibnamefont {Weiss}},\ }\bibfield
  {title} {\enquote {\bibinfo {title} {Spatial solitons in a semiconductor
  microresonator},}\ }\href@noop {} {\bibfield  {journal} {\bibinfo  {journal}
  {Applied Physics B}\ }\textbf {\bibinfo {volume} {72}},\ \bibinfo {pages}
  {377--380} (\bibinfo {year} {2001})}\BibitemShut {NoStop}%
\bibitem [{\citenamefont {Barland}\ \emph {et~al.}(2002)\citenamefont
  {Barland}, \citenamefont {Tredicce}, \citenamefont {Brambilla}, \citenamefont
  {Lugiato}, \citenamefont {Balle}, \citenamefont {Giudici}, \citenamefont
  {Maggipinto}, \citenamefont {Spinelli}, \citenamefont {Tissoni},
  \citenamefont {Knoedl}, \citenamefont {Miller},\ and\ \citenamefont
  {Jaeger}}]{Brambilla3}%
  \BibitemOpen
  \bibfield  {author} {\bibinfo {author} {\bibfnamefont {S.}~\bibnamefont
  {Barland}}, \bibinfo {author} {\bibfnamefont {J.~R.}\ \bibnamefont
  {Tredicce}}, \bibinfo {author} {\bibfnamefont {M.}~\bibnamefont {Brambilla}},
  \bibinfo {author} {\bibfnamefont {L.~A.}\ \bibnamefont {Lugiato}}, \bibinfo
  {author} {\bibfnamefont {S.}~\bibnamefont {Balle}}, \bibinfo {author}
  {\bibfnamefont {M.}~\bibnamefont {Giudici}}, \bibinfo {author} {\bibfnamefont
  {T.}~\bibnamefont {Maggipinto}}, \bibinfo {author} {\bibfnamefont
  {L.}~\bibnamefont {Spinelli}}, \bibinfo {author} {\bibfnamefont
  {G.}~\bibnamefont {Tissoni}}, \bibinfo {author} {\bibfnamefont
  {T.}~\bibnamefont {Knoedl}}, \bibinfo {author} {\bibfnamefont
  {M.}~\bibnamefont {Miller}}, \ and\ \bibinfo {author} {\bibfnamefont
  {R.}~\bibnamefont {Jaeger}},\ }\bibfield  {title} {\enquote {\bibinfo {title}
  {{Cavity solitons as pixels in semiconductor microcavities}},}\ }\href@noop
  {} {\bibfield  {journal} {\bibinfo  {journal} {Nature}\ }\textbf {\bibinfo
  {volume} {419}},\ \bibinfo {pages} {699--702} (\bibinfo {year}
  {2002})}\BibitemShut {NoStop}%
\bibitem [{\citenamefont {Silberberg}(1990)}]{Silberberg1}%
  \BibitemOpen
  \bibfield  {author} {\bibinfo {author} {\bibfnamefont {Y.}~\bibnamefont
  {Silberberg}},\ }\bibfield  {title} {\enquote {\bibinfo {title} {Collapse of
  optical pulses},}\ }\href@noop {} {\bibfield  {journal} {\bibinfo  {journal}
  {Opt. Lett.}\ }\textbf {\bibinfo {volume} {15}},\ \bibinfo {pages} {1282}
  (\bibinfo {year} {1990})}\BibitemShut {NoStop}%
\bibitem [{\citenamefont {Edmundson}(1997)}]{Edmundson1}%
  \BibitemOpen
  \bibfield  {author} {\bibinfo {author} {\bibfnamefont {D.~E.}\ \bibnamefont
  {Edmundson}},\ }\bibfield  {title} {\enquote {\bibinfo {title} {Unstable
  higher modes of three-dimensional nonlinear {S}chr{\"{o}}dinger equation},}\
  }\href@noop {} {\bibfield  {journal} {\bibinfo  {journal} {Phys. Rev. E}\
  }\textbf {\bibinfo {volume} {55}},\ \bibinfo {pages} {7636} (\bibinfo {year}
  {1997})}\BibitemShut {NoStop}%
\bibitem [{\citenamefont {Tlidi}\ \emph
  {et~al.}(1998{\natexlab{a}})\citenamefont {Tlidi}, \citenamefont
  {Haelterman},\ and\ \citenamefont {Mandel}}]{Tlidi4}%
  \BibitemOpen
  \bibfield  {author} {\bibinfo {author} {\bibfnamefont {M.}~\bibnamefont
  {Tlidi}}, \bibinfo {author} {\bibfnamefont {M.}~\bibnamefont {Haelterman}}, \
  and\ \bibinfo {author} {\bibfnamefont {P.}~\bibnamefont {Mandel}},\
  }\bibfield  {title} {\enquote {\bibinfo {title} {{3D} patterns and pattern
  selection in optical bistability},}\ }\href@noop {} {\bibfield  {journal}
  {\bibinfo  {journal} {Europhys. Lett.}\ }\textbf {\bibinfo {volume} {42}},\
  \bibinfo {pages} {505--509} (\bibinfo {year}
  {1998}{\natexlab{a}})}\BibitemShut {NoStop}%
\bibitem [{\citenamefont {Tlidi}\ \emph
  {et~al.}(1998{\natexlab{b}})\citenamefont {Tlidi}, \citenamefont
  {Haelterman},\ and\ \citenamefont {Mandel}}]{Tlidi2}%
  \BibitemOpen
  \bibfield  {author} {\bibinfo {author} {\bibfnamefont {M.}~\bibnamefont
  {Tlidi}}, \bibinfo {author} {\bibfnamefont {M.}~\bibnamefont {Haelterman}}, \
  and\ \bibinfo {author} {\bibfnamefont {P.}~\bibnamefont {Mandel}},\
  }\bibfield  {title} {\enquote {\bibinfo {title} {Three-dimensional structures
  in diffractive and dispersive nonlinear ring cavities},}\ }\href@noop {}
  {\bibfield  {journal} {\bibinfo  {journal} {Quantum Semiclass. Opt.}\
  }\textbf {\bibinfo {volume} {10}},\ \bibinfo {pages} {869--878} (\bibinfo
  {year} {1998}{\natexlab{b}})}\BibitemShut {NoStop}%
\bibitem [{\citenamefont {Brambilla}\ \emph {et~al.}(2004)\citenamefont
  {Brambilla}, \citenamefont {Maggipinto}, \citenamefont {Patera},\ and\
  \citenamefont {Columbo}}]{Brambilla1}%
  \BibitemOpen
  \bibfield  {author} {\bibinfo {author} {\bibfnamefont {M.}~\bibnamefont
  {Brambilla}}, \bibinfo {author} {\bibfnamefont {T.}~\bibnamefont
  {Maggipinto}}, \bibinfo {author} {\bibfnamefont {G.}~\bibnamefont {Patera}},
  \ and\ \bibinfo {author} {\bibfnamefont {L.}~\bibnamefont {Columbo}},\
  }\bibfield  {title} {\enquote {\bibinfo {title} {Cavity light bullets:
  {T}hree-dimensional localized structures in a nonlinear optical resonator},}\
  }\href@noop {} {\bibfield  {journal} {\bibinfo  {journal} {Phys.\ Rev.
  Lett.}\ }\textbf {\bibinfo {volume} {93}},\ \bibinfo {pages} {203901}
  (\bibinfo {year} {2004})}\BibitemShut {NoStop}%
\bibitem [{\citenamefont {Kaliteevski\u{i}}\ and\ \citenamefont
  {Rozanov}(2000)}]{Rozanov1}%
  \BibitemOpen
  \bibfield  {author} {\bibinfo {author} {\bibfnamefont {N.~A.}\ \bibnamefont
  {Kaliteevski\u{i}}}\ and\ \bibinfo {author} {\bibfnamefont {N.~N.}\
  \bibnamefont {Rozanov}},\ }\bibfield  {title} {\enquote {\bibinfo {title} {On
  three-dimensional dissipative optical solitons: Collisions of laser bullets
  and topological solitons},}\ }\href@noop {} {\bibfield  {journal} {\bibinfo
  {journal} {Opt. Spect.}\ }\textbf {\bibinfo {volume} {89}},\ \bibinfo {pages}
  {569--573} (\bibinfo {year} {2000})}\BibitemShut {NoStop}%
\bibitem [{\citenamefont {Veretenov}\ \emph {et~al.}(2000)\citenamefont
  {Veretenov}, \citenamefont {Vladimirov}, \citenamefont {Kaliteevski\v{i}},
  \citenamefont {Rozanov}, \citenamefont {Fedorov},\ and\ \citenamefont
  {Shatsev}}]{Veretenov1}%
  \BibitemOpen
  \bibfield  {author} {\bibinfo {author} {\bibfnamefont {N.~A.}\ \bibnamefont
  {Veretenov}}, \bibinfo {author} {\bibfnamefont {A.~G.}\ \bibnamefont
  {Vladimirov}}, \bibinfo {author} {\bibfnamefont {N.~A.}\ \bibnamefont
  {Kaliteevski\v{i}}}, \bibinfo {author} {\bibfnamefont {N.~N.}\ \bibnamefont
  {Rozanov}}, \bibinfo {author} {\bibfnamefont {S.~V.}\ \bibnamefont
  {Fedorov}}, \ and\ \bibinfo {author} {\bibfnamefont {A.~N.}\ \bibnamefont
  {Shatsev}},\ }\bibfield  {title} {\enquote {\bibinfo {title} {Conditions for
  the existence of laser bullets},}\ }\href@noop {} {\bibfield  {journal}
  {\bibinfo  {journal} {Opt. Spectrosc.}\ }\textbf {\bibinfo {volume} {89}},\
  \bibinfo {pages} {380} (\bibinfo {year} {2000})}\BibitemShut {NoStop}%
\bibitem [{\citenamefont {Marconi}\ \emph {et~al.}(2014)\citenamefont
  {Marconi}, \citenamefont {Javaloyes}, \citenamefont {Balle},\ and\
  \citenamefont {Giudici}}]{Marconi1}%
  \BibitemOpen
  \bibfield  {author} {\bibinfo {author} {\bibfnamefont {M.}~\bibnamefont
  {Marconi}}, \bibinfo {author} {\bibfnamefont {J.}~\bibnamefont {Javaloyes}},
  \bibinfo {author} {\bibfnamefont {S.}~\bibnamefont {Balle}}, \ and\ \bibinfo
  {author} {\bibfnamefont {M.}~\bibnamefont {Giudici}},\ }\bibfield  {title}
  {\enquote {\bibinfo {title} {How lasing localized structures evolve out of
  passive mode locking},}\ }\href@noop {} {\bibfield  {journal} {\bibinfo
  {journal} {Phys. Rev. Lett.}\ }\textbf {\bibinfo {volume} {112}},\ \bibinfo
  {pages} {223901} (\bibinfo {year} {2014})}\BibitemShut {NoStop}%
\bibitem [{\citenamefont {Javaloyes}(2016)}]{Javaloyes1}%
  \BibitemOpen
  \bibfield  {author} {\bibinfo {author} {\bibfnamefont {J.}~\bibnamefont
  {Javaloyes}},\ }\bibfield  {title} {\enquote {\bibinfo {title} {Cavity light
  bullets in passively mode-locked semiconductor lasers},}\ }\href@noop {}
  {\bibfield  {journal} {\bibinfo  {journal} {Phys. Rev. Lett.}\ }\textbf
  {\bibinfo {volume} {116}},\ \bibinfo {pages} {043901} (\bibinfo {year}
  {2016})}\BibitemShut {NoStop}%
\bibitem [{\citenamefont {Dohmen}\ \emph {et~al.}(2020)\citenamefont {Dohmen},
  \citenamefont {Javaloyes},\ and\ \citenamefont {Gurevich}}]{Gurev1}%
  \BibitemOpen
  \bibfield  {author} {\bibinfo {author} {\bibfnamefont {F.}~\bibnamefont
  {Dohmen}}, \bibinfo {author} {\bibfnamefont {J.}~\bibnamefont {Javaloyes}}, \
  and\ \bibinfo {author} {\bibfnamefont {S.~V.}\ \bibnamefont {Gurevich}},\
  }\bibfield  {title} {\enquote {\bibinfo {title} {Bound states of light
  bullets in passively mode-locked semiconductor lasers},}\ }\href@noop {}
  {\bibfield  {journal} {\bibinfo  {journal} {Chaos}\ }\textbf {\bibinfo
  {volume} {30}},\ \bibinfo {pages} {063120} (\bibinfo {year}
  {2020})}\BibitemShut {NoStop}%
\bibitem [{\citenamefont {Staliunas}(1998)}]{Staliunas1}%
  \BibitemOpen
  \bibfield  {author} {\bibinfo {author} {\bibfnamefont {K.}~\bibnamefont
  {Staliunas}},\ }\bibfield  {title} {\enquote {\bibinfo {title}
  {Three-dimensional {T}uring structures and spatial solitons in optical
  parametric oscillators},}\ }\href@noop {} {\bibfield  {journal} {\bibinfo
  {journal} {Phys. Rev. Lett.}\ }\textbf {\bibinfo {volume} {81}},\ \bibinfo
  {pages} {81--84} (\bibinfo {year} {1998})}\BibitemShut {NoStop}%
\bibitem [{\citenamefont {Veretenov}\ and\ \citenamefont
  {Tlidi}(2009)}]{Veretenov2}%
  \BibitemOpen
  \bibfield  {author} {\bibinfo {author} {\bibfnamefont {N.~A.}\ \bibnamefont
  {Veretenov}}\ and\ \bibinfo {author} {\bibfnamefont {M.}~\bibnamefont
  {Tlidi}},\ }\bibfield  {title} {\enquote {\bibinfo {title} {Dissipative light
  bullets in an optical parametric oscillator},}\ }\href@noop {} {\bibfield
  {journal} {\bibinfo  {journal} {Phys. Rev. A}\ }\textbf {\bibinfo {volume}
  {80}},\ \bibinfo {pages} {023822} (\bibinfo {year} {2009})}\BibitemShut
  {NoStop}%
\bibitem [{\citenamefont {Panoiu}\ \emph {et~al.}(2005)\citenamefont {Panoiu},
  \citenamefont {Osgood{, Jr.}}, \citenamefont {Malomed}, \citenamefont
  {Lederer}, \citenamefont {Mazilu},\ and\ \citenamefont
  {Mihalache}}]{Panoiu1}%
  \BibitemOpen
  \bibfield  {author} {\bibinfo {author} {\bibfnamefont {N.~{-C}.}\
  \bibnamefont {Panoiu}}, \bibinfo {author} {\bibfnamefont {R.~M.}\
  \bibnamefont {Osgood{, Jr.}}}, \bibinfo {author} {\bibfnamefont {B.~A.}\
  \bibnamefont {Malomed}}, \bibinfo {author} {\bibfnamefont {F.}~\bibnamefont
  {Lederer}}, \bibinfo {author} {\bibfnamefont {D.}~\bibnamefont {Mazilu}}, \
  and\ \bibinfo {author} {\bibfnamefont {D.}~\bibnamefont {Mihalache}},\
  }\bibfield  {title} {\enquote {\bibinfo {title} {Parametric light bullets
  supported by quasi-phase-matched quadratically nonlinear crystals},}\
  }\href@noop {} {\bibfield  {journal} {\bibinfo  {journal} {Phys. Rev. E}\
  }\textbf {\bibinfo {volume} {71}},\ \bibinfo {pages} {036615} (\bibinfo
  {year} {2005})}\BibitemShut {NoStop}%
\bibitem [{\citenamefont {Tlidi}\ and\ \citenamefont {Mandel}(1999)}]{Tlidi3}%
  \BibitemOpen
  \bibfield  {author} {\bibinfo {author} {\bibfnamefont {M.}~\bibnamefont
  {Tlidi}}\ and\ \bibinfo {author} {\bibfnamefont {P.}~\bibnamefont {Mandel}},\
  }\bibfield  {title} {\enquote {\bibinfo {title} {Three-dimensional optical
  crystals and localized structures in cavity second harmonic generation},}\
  }\href@noop {} {\bibfield  {journal} {\bibinfo  {journal} {Phys.\ Rev.
  Lett.}\ }\textbf {\bibinfo {volume} {83}},\ \bibinfo {pages} {4995} (\bibinfo
  {year} {1999})}\BibitemShut {NoStop}%
\bibitem [{\citenamefont {Tlidi}(2000)}]{tlidi2000three}%
  \BibitemOpen
  \bibfield  {author} {\bibinfo {author} {\bibfnamefont {Mustapha}\
  \bibnamefont {Tlidi}},\ }\bibfield  {title} {\enquote {\bibinfo {title}
  {Three-dimensional crystals and localized structures in diffractive and
  dispersive nonlinear ring cavities},}\ }\href@noop {} {\bibfield  {journal}
  {\bibinfo  {journal} {Journal of Optics B: Quantum and Semiclassical Optics}\
  }\textbf {\bibinfo {volume} {2}},\ \bibinfo {pages} {438} (\bibinfo {year}
  {2000})}\BibitemShut {NoStop}%
\bibitem [{\citenamefont {Grelu}\ and\ \citenamefont
  {Akhmediev}(2012)}]{Grelu12}%
  \BibitemOpen
  \bibfield  {author} {\bibinfo {author} {\bibfnamefont {P.}~\bibnamefont
  {Grelu}}\ and\ \bibinfo {author} {\bibfnamefont {N.}~\bibnamefont
  {Akhmediev}},\ }\bibfield  {title} {\enquote {\bibinfo {title} {Dissipative
  solitons for mode-locked lasers.}}\ }\href@noop {} {\bibfield  {journal}
  {\bibinfo  {journal} {Nat. Photonics}\ }\textbf {\bibinfo {volume} {6}},\
  \bibinfo {pages} {84} (\bibinfo {year} {2012})}\BibitemShut {NoStop}%
\bibitem [{\citenamefont {Kockaert}\ \emph {et~al.}(2006)\citenamefont
  {Kockaert}, \citenamefont {Tassin}, \citenamefont {Van~der Sande},
  \citenamefont {Veretennicoff},\ and\ \citenamefont
  {Tlidi}}]{kockaert2006negative}%
  \BibitemOpen
  \bibfield  {author} {\bibinfo {author} {\bibfnamefont {Pascal}\ \bibnamefont
  {Kockaert}}, \bibinfo {author} {\bibfnamefont {Philippe}\ \bibnamefont
  {Tassin}}, \bibinfo {author} {\bibfnamefont {Guy}\ \bibnamefont {Van~der
  Sande}}, \bibinfo {author} {\bibfnamefont {Irina}\ \bibnamefont
  {Veretennicoff}}, \ and\ \bibinfo {author} {\bibfnamefont {Mustapha}\
  \bibnamefont {Tlidi}},\ }\bibfield  {title} {\enquote {\bibinfo {title}
  {Negative diffraction pattern dynamics in nonlinear cavities with left-handed
  materials},}\ }\href@noop {} {\bibfield  {journal} {\bibinfo  {journal}
  {Physical Review A}\ }\textbf {\bibinfo {volume} {74}},\ \bibinfo {pages}
  {033822} (\bibinfo {year} {2006})}\BibitemShut {NoStop}%
\bibitem [{\citenamefont {Mili{\'{a}}n}\ \emph {et~al.}(2019)\citenamefont
  {Mili{\'{a}}n}, \citenamefont {Kartashov},\ and\ \citenamefont
  {Torner}}]{Torner1}%
  \BibitemOpen
  \bibfield  {author} {\bibinfo {author} {\bibfnamefont {C.}~\bibnamefont
  {Mili{\'{a}}n}}, \bibinfo {author} {\bibfnamefont {Y.~V.}\ \bibnamefont
  {Kartashov}}, \ and\ \bibinfo {author} {\bibfnamefont {L.}~\bibnamefont
  {Torner}},\ }\bibfield  {title} {\enquote {\bibinfo {title} {Robust
  ultrashort light bullets in strongly twisted waveguide arrays},}\ }\href@noop
  {} {\bibfield  {journal} {\bibinfo  {journal} {Phys.\ Rev. Lett.}\ }\textbf
  {\bibinfo {volume} {123}},\ \bibinfo {pages} {133902} (\bibinfo {year}
  {2019})}\BibitemShut {NoStop}%
\bibitem [{\citenamefont {Bordeu}\ and\ \citenamefont {Clerc}(2015)}]{Clerc}%
  \BibitemOpen
  \bibfield  {author} {\bibinfo {author} {\bibfnamefont {Ignacio}\ \bibnamefont
  {Bordeu}}\ and\ \bibinfo {author} {\bibfnamefont {Marcel~G.}\ \bibnamefont
  {Clerc}},\ }\bibfield  {title} {\enquote {\bibinfo {title} {Rodlike localized
  structure in isotropic pattern-forming systems},}\ }\href@noop {} {\bibfield
  {journal} {\bibinfo  {journal} {Phys. Rev. E}\ }\textbf {\bibinfo {volume}
  {92}},\ \bibinfo {pages} {042915} (\bibinfo {year} {2015})}\BibitemShut
  {NoStop}%
\bibitem [{\citenamefont {Mihalache}\ \emph {et~al.}(2006)\citenamefont
  {Mihalache}, \citenamefont {Mazilu}, \citenamefont {Lederer}, \citenamefont
  {Kartashov}, \citenamefont {Crasovan}, \citenamefont {Torner},\ and\
  \citenamefont {Malomed}}]{mihalache2006stable}%
  \BibitemOpen
  \bibfield  {author} {\bibinfo {author} {\bibfnamefont {Dumitru}\ \bibnamefont
  {Mihalache}}, \bibinfo {author} {\bibfnamefont {D}~\bibnamefont {Mazilu}},
  \bibinfo {author} {\bibfnamefont {F}~\bibnamefont {Lederer}}, \bibinfo
  {author} {\bibfnamefont {Yaroslav~V}\ \bibnamefont {Kartashov}}, \bibinfo
  {author} {\bibfnamefont {L-C}\ \bibnamefont {Crasovan}}, \bibinfo {author}
  {\bibfnamefont {L}~\bibnamefont {Torner}}, \ and\ \bibinfo {author}
  {\bibfnamefont {BA}~\bibnamefont {Malomed}},\ }\bibfield  {title} {\enquote
  {\bibinfo {title} {Stable vortex tori in the three-dimensional cubic-quintic
  ginzburg-landau equation},}\ }\href@noop {} {\bibfield  {journal} {\bibinfo
  {journal} {Physical Review Letters}\ }\textbf {\bibinfo {volume} {97}},\
  \bibinfo {pages} {073904} (\bibinfo {year} {2006})}\BibitemShut {NoStop}%
\bibitem [{\citenamefont {Malomed}\ \emph {et~al.}(2005)\citenamefont
  {Malomed}, \citenamefont {Mihalache}, \citenamefont {Wise},\ and\
  \citenamefont {Torner}}]{Malomed1}%
  \BibitemOpen
  \bibfield  {author} {\bibinfo {author} {\bibfnamefont {B.~A.}\ \bibnamefont
  {Malomed}}, \bibinfo {author} {\bibfnamefont {D.}~\bibnamefont {Mihalache}},
  \bibinfo {author} {\bibfnamefont {F.}~\bibnamefont {Wise}}, \ and\ \bibinfo
  {author} {\bibfnamefont {L.}~\bibnamefont {Torner}},\ }\bibfield  {title}
  {\enquote {\bibinfo {title} {Spatiotemporal optical solitons},}\ }\href@noop
  {} {\bibfield  {journal} {\bibinfo  {journal} {J. Opt. B: Quantum Semiclass.
  Opt.}\ }\textbf {\bibinfo {volume} {7}},\ \bibinfo {pages} {R53--R72}
  (\bibinfo {year} {2005})}\BibitemShut {NoStop}%
\bibitem [{\citenamefont {Mihalache}(2014)}]{Mihalache14}%
  \BibitemOpen
  \bibfield  {author} {\bibinfo {author} {\bibfnamefont {D.}~\bibnamefont
  {Mihalache}},\ }\bibfield  {title} {\enquote {\bibinfo {title}
  {Multidimensional localized stuctures in optics and {B}ose--{E}instein
  condensates: a selection of recent studies.}}\ }\href@noop {} {\bibfield
  {journal} {\bibinfo  {journal} {Rom. J. Phys.}\ }\textbf {\bibinfo {volume}
  {59}},\ \bibinfo {pages} {295} (\bibinfo {year} {2014})}\BibitemShut
  {NoStop}%
\bibitem [{\citenamefont {Malomed}\ and\ \citenamefont
  {Mihalache}(2019)}]{Malomed19}%
  \BibitemOpen
  \bibfield  {author} {\bibinfo {author} {\bibfnamefont {B.~A.}\ \bibnamefont
  {Malomed}}\ and\ \bibinfo {author} {\bibfnamefont {D.}~\bibnamefont
  {Mihalache}},\ }\bibfield  {title} {\enquote {\bibinfo {title} {Nonlinear
  waves in optical and matter-wave media: a topical survey of recent
  theoretical and experimental results.}}\ }\href@noop {} {\bibfield  {journal}
  {\bibinfo  {journal} {Rom. J. Phys.}\ }\textbf {\bibinfo {volume} {64}},\
  \bibinfo {pages} {106} (\bibinfo {year} {2019})}\BibitemShut {NoStop}%
\bibitem [{\citenamefont {Lugiato}\ and\ \citenamefont
  {Lefever}(1987)}]{Lefever1}%
  \BibitemOpen
  \bibfield  {author} {\bibinfo {author} {\bibfnamefont {L.~A.}\ \bibnamefont
  {Lugiato}}\ and\ \bibinfo {author} {\bibfnamefont {R.}~\bibnamefont
  {Lefever}},\ }\bibfield  {title} {\enquote {\bibinfo {title} {Spatial
  dissipative structures in passive optical systems},}\ }\href@noop {}
  {\bibfield  {journal} {\bibinfo  {journal} {Phys. Rev. Lett.}\ }\textbf
  {\bibinfo {volume} {58}},\ \bibinfo {pages} {2209} (\bibinfo {year}
  {1987})}\BibitemShut {NoStop}%
\bibitem [{\citenamefont {Haelterman}\ \emph {et~al.}(1992)\citenamefont
  {Haelterman}, \citenamefont {Trillo},\ and\ \citenamefont
  {Wabnitz}}]{haelterman1992dissipative}%
  \BibitemOpen
  \bibfield  {author} {\bibinfo {author} {\bibfnamefont {Marc}\ \bibnamefont
  {Haelterman}}, \bibinfo {author} {\bibfnamefont {Stefano}\ \bibnamefont
  {Trillo}}, \ and\ \bibinfo {author} {\bibfnamefont {Stefan}\ \bibnamefont
  {Wabnitz}},\ }\bibfield  {title} {\enquote {\bibinfo {title} {Dissipative
  modulation instability in a nonlinear dispersive ring cavity},}\ }\href@noop
  {} {\bibfield  {journal} {\bibinfo  {journal} {Optics communications}\
  }\textbf {\bibinfo {volume} {91}},\ \bibinfo {pages} {401--407} (\bibinfo
  {year} {1992})}\BibitemShut {NoStop}%
\bibitem [{\citenamefont {Peschel}\ \emph {et~al.}(2004)\citenamefont
  {Peschel}, \citenamefont {Egorov},\ and\ \citenamefont
  {Lederer}}]{peschel2004discrete}%
  \BibitemOpen
  \bibfield  {author} {\bibinfo {author} {\bibfnamefont {U}~\bibnamefont
  {Peschel}}, \bibinfo {author} {\bibfnamefont {O}~\bibnamefont {Egorov}}, \
  and\ \bibinfo {author} {\bibfnamefont {F}~\bibnamefont {Lederer}},\
  }\bibfield  {title} {\enquote {\bibinfo {title} {Discrete cavity solitons},}\
  }in\ \href@noop {} {\emph {\bibinfo {booktitle} {Nonlinear Guided Waves and
  Their Applications}}}\ (\bibinfo {organization} {Optical Society of
  America},\ \bibinfo {year} {2004})\ p.\ \bibinfo {pages} {WB6}\BibitemShut
  {NoStop}%
\bibitem [{\citenamefont {Chembo}\ and\ \citenamefont
  {Menyuk}(2013)}]{Chembo1}%
  \BibitemOpen
  \bibfield  {author} {\bibinfo {author} {\bibfnamefont {Y.~K.}\ \bibnamefont
  {Chembo}}\ and\ \bibinfo {author} {\bibfnamefont {C.~R.}\ \bibnamefont
  {Menyuk}},\ }\bibfield  {title} {\enquote {\bibinfo {title} {Spatiotemporal
  {L}ugiato--{L}efever formalism for {K}err-comb generation in
  whispering-gallery-mode resonators},}\ }\href@noop {} {\bibfield  {journal}
  {\bibinfo  {journal} {Phys. Rev. A}\ }\textbf {\bibinfo {volume} {87}},\
  \bibinfo {pages} {053852} (\bibinfo {year} {2013})}\BibitemShut {NoStop}%
\bibitem [{\citenamefont {Morales}\ and\ \citenamefont
  {Lee}(1974)}]{morales1974ponderomotive}%
  \BibitemOpen
  \bibfield  {author} {\bibinfo {author} {\bibfnamefont {GJ}~\bibnamefont
  {Morales}}\ and\ \bibinfo {author} {\bibfnamefont {YC}~\bibnamefont {Lee}},\
  }\bibfield  {title} {\enquote {\bibinfo {title} {Ponderomotive-force effects
  in a nonuniform plasma},}\ }\href@noop {} {\bibfield  {journal} {\bibinfo
  {journal} {Physical Review Letters}\ }\textbf {\bibinfo {volume} {33}},\
  \bibinfo {pages} {1016} (\bibinfo {year} {1974})}\BibitemShut {NoStop}%
\bibitem [{\citenamefont {Kaup}\ and\ \citenamefont
  {Newell}(1978)}]{kaup1978theory}%
  \BibitemOpen
  \bibfield  {author} {\bibinfo {author} {\bibfnamefont {David~James}\
  \bibnamefont {Kaup}}\ and\ \bibinfo {author} {\bibfnamefont {Alan~C}\
  \bibnamefont {Newell}},\ }\bibfield  {title} {\enquote {\bibinfo {title}
  {Theory of nonlinear oscillating dipolar excitations in one-dimensional
  condensates},}\ }\href@noop {} {\bibfield  {journal} {\bibinfo  {journal}
  {Physical Review B}\ }\textbf {\bibinfo {volume} {18}},\ \bibinfo {pages}
  {5162} (\bibinfo {year} {1978})}\BibitemShut {NoStop}%
\bibitem [{\citenamefont {Chembo}\ \emph {et~al.}(2017)\citenamefont {Chembo},
  \citenamefont {Gomila}, \citenamefont {Tlidi},\ and\ \citenamefont
  {Menyuk}}]{Chembo2017theory}%
  \BibitemOpen
  \bibfield  {author} {\bibinfo {author} {\bibfnamefont {Yanne~K}\ \bibnamefont
  {Chembo}}, \bibinfo {author} {\bibfnamefont {Dami{\`a}}\ \bibnamefont
  {Gomila}}, \bibinfo {author} {\bibfnamefont {Mustapha}\ \bibnamefont
  {Tlidi}}, \ and\ \bibinfo {author} {\bibfnamefont {Curtis~R}\ \bibnamefont
  {Menyuk}},\ }\href@noop {} {\enquote {\bibinfo {title} {Theory and
  applications of the {L}ugiato--{L}efever equation},}\ } (\bibinfo {year}
  {2017})\BibitemShut {NoStop}%
\bibitem [{\citenamefont {Firth}\ \emph {et~al.}(1992)\citenamefont {Firth},
  \citenamefont {Scroggie}, \citenamefont {McDonald},\ and\ \citenamefont
  {Lugiato}}]{PhysRevA.46.R3609}%
  \BibitemOpen
  \bibfield  {author} {\bibinfo {author} {\bibfnamefont {W.~J.}\ \bibnamefont
  {Firth}}, \bibinfo {author} {\bibfnamefont {A.~J.}\ \bibnamefont {Scroggie}},
  \bibinfo {author} {\bibfnamefont {G.~S.}\ \bibnamefont {McDonald}}, \ and\
  \bibinfo {author} {\bibfnamefont {L.~A.}\ \bibnamefont {Lugiato}},\
  }\bibfield  {title} {\enquote {\bibinfo {title} {Hexagonal patterns in
  optical bistability},}\ }\href@noop {} {\bibfield  {journal} {\bibinfo
  {journal} {Phys. Rev. A}\ }\textbf {\bibinfo {volume} {46}},\ \bibinfo
  {pages} {R3609--R3612} (\bibinfo {year} {1992})}\BibitemShut {NoStop}%
\bibitem [{\citenamefont {Tlidi}\ \emph {et~al.}(1996)\citenamefont {Tlidi},
  \citenamefont {Lefever},\ and\ \citenamefont {Mandel}}]{Tlidi6}%
  \BibitemOpen
  \bibfield  {author} {\bibinfo {author} {\bibfnamefont {M.}~\bibnamefont
  {Tlidi}}, \bibinfo {author} {\bibfnamefont {R.}~\bibnamefont {Lefever}}, \
  and\ \bibinfo {author} {\bibfnamefont {P.}~\bibnamefont {Mandel}},\
  }\bibfield  {title} {\enquote {\bibinfo {title} {Pattern selection in optical
  bistability},}\ }\href@noop {} {\bibfield  {journal} {\bibinfo  {journal}
  {Quantum Semicalss. Opt.}\ }\textbf {\bibinfo {volume} {8}},\ \bibinfo
  {pages} {931--938} (\bibinfo {year} {1996})}\BibitemShut {NoStop}%
\bibitem [{\citenamefont {Jones}\ and\ \citenamefont
  {O'{B}rian}(1996)}]{Jones1}%
  \BibitemOpen
  \bibfield  {author} {\bibinfo {author} {\bibfnamefont {W.~B.}\ \bibnamefont
  {Jones}}\ and\ \bibinfo {author} {\bibfnamefont {J.}~\bibnamefont
  {O'{B}rian}},\ }\bibfield  {title} {\enquote {\bibinfo {title}
  {Pseudo-spectral methods and linear instabilities in reaction-diffusion
  fronts},}\ }\href@noop {} {\bibfield  {journal} {\bibinfo  {journal} {Chaos}\
  }\textbf {\bibinfo {volume} {6}},\ \bibinfo {pages} {219--228} (\bibinfo
  {year} {1996})}\BibitemShut {NoStop}%
\bibitem [{\citenamefont {Kassam}(2004)}]{Kassam2}%
  \BibitemOpen
  \bibfield  {author} {\bibinfo {author} {\bibfnamefont {A.{-K}.}\ \bibnamefont
  {Kassam}},\ }\href@noop {} {\emph {\bibinfo {title} {High order timestepping
  for stiff semilinear partial differential equations}}}\ (\bibinfo
  {publisher} {University of {O}xford},\ \bibinfo {year} {2004})\BibitemShut
  {NoStop}%
\bibitem [{\citenamefont {Trefethen}(2000)}]{Trefethen1}%
  \BibitemOpen
  \bibfield  {author} {\bibinfo {author} {\bibfnamefont {L.~N.}\ \bibnamefont
  {Trefethen}},\ }\href@noop {} {\emph {\bibinfo {title} {Spectral {M}ethods in
  {MATLAB}}}}\ (\bibinfo  {publisher} {{SIAM}, Philadelphia},\ \bibinfo {year}
  {2000})\BibitemShut {NoStop}%
\bibitem [{\citenamefont {Cox}\ and\ \citenamefont
  {Matthews}(2002)}]{Matthews1}%
  \BibitemOpen
  \bibfield  {author} {\bibinfo {author} {\bibfnamefont {S.~M.}\ \bibnamefont
  {Cox}}\ and\ \bibinfo {author} {\bibfnamefont {P.~C.}\ \bibnamefont
  {Matthews}},\ }\bibfield  {title} {\enquote {\bibinfo {title} {Exponential
  time differencing for stiff systems},}\ }\href@noop {} {\bibfield  {journal}
  {\bibinfo  {journal} {J. Comput. Phys.}\ }\textbf {\bibinfo {volume} {176}},\
  \bibinfo {pages} {430--455} (\bibinfo {year} {2002})}\BibitemShut {NoStop}%
\bibitem [{\citenamefont {Saad}(2011)}]{Saad1}%
  \BibitemOpen
  \bibfield  {author} {\bibinfo {author} {\bibfnamefont {Y.}~\bibnamefont
  {Saad}},\ }\href@noop {} {\emph {\bibinfo {title} {Numerical methods for
  large eigenvalue problems}}}\ (\bibinfo  {publisher} {{SIAM}},\ \bibinfo
  {year} {2011})\BibitemShut {NoStop}%
\bibitem [{\citenamefont {Gopalakrishnan}\ \emph {et~al.}(2021)\citenamefont
  {Gopalakrishnan}, \citenamefont {Panajotov}, \citenamefont {Taki},\ and\
  \citenamefont {Tlidi}}]{gopalakrishnan2021dissipative}%
  \BibitemOpen
  \bibfield  {author} {\bibinfo {author} {\bibfnamefont {Shyam~Sunder}\
  \bibnamefont {Gopalakrishnan}}, \bibinfo {author} {\bibfnamefont {Krassimir}\
  \bibnamefont {Panajotov}}, \bibinfo {author} {\bibfnamefont {Majid}\
  \bibnamefont {Taki}}, \ and\ \bibinfo {author} {\bibfnamefont {Mustapha}\
  \bibnamefont {Tlidi}},\ }\bibfield  {title} {\enquote {\bibinfo {title}
  {Dissipative light bullets in kerr cavities: Multistability, clustering, and
  rogue waves},}\ }\href@noop {} {\bibfield  {journal} {\bibinfo  {journal}
  {Physical review letters}\ }\textbf {\bibinfo {volume} {126}},\ \bibinfo
  {pages} {153902} (\bibinfo {year} {2021})}\BibitemShut {NoStop}%
\bibitem [{\citenamefont {Gomila}\ \emph {et~al.}(2007)\citenamefont {Gomila},
  \citenamefont {Scroggie},\ and\ \citenamefont
  {Firth}}]{gomila2007bifurcation}%
  \BibitemOpen
  \bibfield  {author} {\bibinfo {author} {\bibfnamefont {Damia}\ \bibnamefont
  {Gomila}}, \bibinfo {author} {\bibfnamefont {Andrew~J}\ \bibnamefont
  {Scroggie}}, \ and\ \bibinfo {author} {\bibfnamefont {William~J}\
  \bibnamefont {Firth}},\ }\bibfield  {title} {\enquote {\bibinfo {title}
  {Bifurcation structure of dissipative solitons},}\ }\href@noop {} {\bibfield
  {journal} {\bibinfo  {journal} {Physica D: Nonlinear Phenomena}\ }\textbf
  {\bibinfo {volume} {227}},\ \bibinfo {pages} {70--77} (\bibinfo {year}
  {2007})}\BibitemShut {NoStop}%
\bibitem [{\citenamefont {Tlidi}\ and\ \citenamefont
  {Gelens}(2010)}]{tlidi2010high}%
  \BibitemOpen
  \bibfield  {author} {\bibinfo {author} {\bibfnamefont {Mustapha}\
  \bibnamefont {Tlidi}}\ and\ \bibinfo {author} {\bibfnamefont {Lendert}\
  \bibnamefont {Gelens}},\ }\bibfield  {title} {\enquote {\bibinfo {title}
  {High-order dispersion stabilizes dark dissipative solitons in all-fiber
  cavities},}\ }\href@noop {} {\bibfield  {journal} {\bibinfo  {journal}
  {Optics letters}\ }\textbf {\bibinfo {volume} {35}},\ \bibinfo {pages}
  {306--308} (\bibinfo {year} {2010})}\BibitemShut {NoStop}%
\bibitem [{\citenamefont {Woods}\ and\ \citenamefont
  {Champneys}(1999)}]{woods_heteroclinic_1999}%
  \BibitemOpen
  \bibfield  {author} {\bibinfo {author} {\bibfnamefont {P.~D}\ \bibnamefont
  {Woods}}\ and\ \bibinfo {author} {\bibfnamefont {A.~R}\ \bibnamefont
  {Champneys}},\ }\bibfield  {title} {\enquote {\bibinfo {title} {Heteroclinic
  tangles and homoclinic snaking in the unfolding of a degenerate reversible
  {Hamiltonian}–{Hopf} bifurcation},}\ }\href@noop {} {\bibfield  {journal}
  {\bibinfo  {journal} {Physica D: Nonlinear Phenomena}\ }\textbf {\bibinfo
  {volume} {129}},\ \bibinfo {pages} {147--170} (\bibinfo {year}
  {1999})}\BibitemShut {NoStop}%
\bibitem [{\citenamefont {Lloyd}\ \emph {et~al.}(2008)\citenamefont {Lloyd},
  \citenamefont {Sandstede}, \citenamefont {Avitabile},\ and\ \citenamefont
  {Champneys}}]{lloyd2008localized}%
  \BibitemOpen
  \bibfield  {author} {\bibinfo {author} {\bibfnamefont {David~JB}\
  \bibnamefont {Lloyd}}, \bibinfo {author} {\bibfnamefont {Bj{\"o}rn}\
  \bibnamefont {Sandstede}}, \bibinfo {author} {\bibfnamefont {Daniele}\
  \bibnamefont {Avitabile}}, \ and\ \bibinfo {author} {\bibfnamefont {Alan~R}\
  \bibnamefont {Champneys}},\ }\bibfield  {title} {\enquote {\bibinfo {title}
  {Localized hexagon patterns of the planar swift--hohenberg equation},}\
  }\href@noop {} {\bibfield  {journal} {\bibinfo  {journal} {SIAM Journal on
  Applied Dynamical Systems}\ }\textbf {\bibinfo {volume} {7}},\ \bibinfo
  {pages} {1049--1100} (\bibinfo {year} {2008})}\BibitemShut {NoStop}%
\bibitem [{\citenamefont {Knobloch}(2015)}]{knobloch2015spatial}%
  \BibitemOpen
  \bibfield  {author} {\bibinfo {author} {\bibfnamefont {E}~\bibnamefont
  {Knobloch}},\ }\bibfield  {title} {\enquote {\bibinfo {title} {Spatial
  localization in dissipative systems},}\ }\href@noop {} {\bibfield  {journal}
  {\bibinfo  {journal} {conmatphys}\ }\textbf {\bibinfo {volume} {6}},\
  \bibinfo {pages} {325--359} (\bibinfo {year} {2015})}\BibitemShut {NoStop}%
\end{thebibliography}%

\end{document}